\documentclass{aa}  

\usepackage{graphicx}
\usepackage{txfonts}
\usepackage{lipsum}
\usepackage{subcaption}         % necessary for continued figures, example in section 3
\usepackage{lscape}             % to rotate a single page table, example in appendix.
\usepackage{placeins}           % useful with \FloatBarrier, to keep 
\usepackage[colorlinks=true,citecolor=blue]{hyperref} 
                                
\begin{document}

%%%%%%%%%%%%%%%%%%%%%%%%%%%%%%%%%%%%%%%%
% if you use custom commands in your title,
% ensure to check your title when submitting!
%%%%%%%%%%%%%%%%%%%%%%%%%%%%%%%%%%%%%%%%
   \title{NOCTURNE. VI. \\ Hour-timescale radio variability in early-stage active galactic nucleus SDSS J122844.81+501751.2}

%   \subtitle{Subtitle}

%%%%%%%%%%%%%%%%%%%%%%%%%%%%%%%%%%%%%%%%
% Please separate each author with the \and command
%
% Use the \corrauth to provide the corresponding
% author address. It will be automatically inserted as 
% footnote in the PDF output.
%
% Please DO NOT include ORCIDs next to author names.
% Instead, please provide an active address for each coauthor:
% it will be automatically extracted by EDPS editorial system, 
% and co-authors will be be able to authenticate their ORCID.
%
% Only authenticated ORCIDs will be taken into account.
% ORCIDs included here will be removed.
%%%%%%%%%%%%%%%%%%%%%%%%%%%%%%%%%%%%%%%%

   \author{E. Järvelä\inst{1}\corrauth{ejarvela@ttu.edu}        % use \corrauth for the corresponding author
        \and P. Romano \inst{2}
        \and J. Nättilä\inst{3}
        \and J.~D. Paul\inst{1}
        \and A. Lähteenmäki\inst{4,5} %\email{fab@nestor-edp.org} %\fnmsep\thanks{NASA fellow (shows the usage of elements in the author field)}
        \and M. Berton\inst{6}%\email{cat@nestor-edp.org}
        \and T. Savolainen\inst{4,5}
        \and S. Panda\inst{7}\thanks{Gemini Science Fellow}
        \and A.~C.~S. Readhead\inst{8}
        \and M. Tornikoski\inst{4}
        }

   \institute{Department of Physics and Astronomy, Texas Tech University, Box 41051, Lubbock, 79409-1051, TX, USA
   \and
   INAF – Osservatorio Astronomico di Brera, Via E. Bianchi 46, I-23807 Merate, Italy
   \and
   Department of Physics, University of Helsinki, P.O. Box 64, FI-00014 University of Helsinki, Finland
   \and
   Metsähovi Radio Observatory, Aalto University, Metsähovintie 114, FI-02540 Kylmälä, Finland
   \and
   Department of Electronics and Nanoengineering, Aalto University, PO Box 15500, FI-00076 Aalto, Finland
   \and
   European Southern Observatory (ESO), Alonso de Córdova 3107, Casilla 19, Santiago 19001, Chile
   \and 
   International Gemini Observatory/NSF NOIRLab, Casilla 603, La Serena, Chile
   \and
   Owens Valley Radio Observatory, California Institute of Technology, Pasadena, CA 91125, USA}

   \date{Received MM DD, 2026}

% \abstract{}{}{}{}{}
% 5 {} token are mandatory
 
  \abstract
   {Recently, a sample of radio-weak active galactic nuclei (AGN) in an early evolutionary stage was discovered to exhibit extreme Jy-level radio flares at 37~GHz. The origin of these extraordinary flares remains unknown. All sources host low-mass supermassive black holes, and most of them can be classified as narrow-line Seyfert 1 galaxies.}
   {We employed multiwavelength observations to investigate one of these sources, SDSS J122844.81+501751.2, during a period of high 37~GHz activity. Our data include radio observations from Metsähovi Radio Observatory (MRO) at 37~GHz, Owens Valley Radio Observatory at 15~GHz, and the Karl G. Jansky Very Large Array (JVLA) at 15, 22, and 33~GHz, as well as optical, ultraviolet, and X-ray observations from \textit{Swift}/XRT and UVOT.} 
   {We studied the variability of the multiwavelength light curves using simple statistical tests, as well as derived the $e$-folding timescales of the flares when possible. Utilising the timescales and the observed flare properties, we estimated the variability brightness temperatures and the variability Doppler factors associated with the flares.}
   {Most individual light curves do not show statistically significant variability, which is likely due to the low number of measurements. The shortest $e$-folding timescale we estimated is only 1.1~hours, measured between MRO 37~GHz and JVLA 33~GHz detections. This is the fastest intrinsic radio variability reported in an AGN so far. This yielded an extreme variability brightness temperature of $T_\mathrm{B}$ = 4.20 $\substack{+1.53 \\ -1.35}~\times$ 10$^{18}$~K and a variability Doppler factor of 438 $\substack{48 \\ -53}$. We discuss magnetic eruptions near the black hole as a possible explanation for the extreme radio flares, and show that both incoherent and coherent processes can produce flares at these timescales.}
  %“Conclusions” can be used to explicitly state the general conclusions that can be drawn from the paper.
  {These extreme radio flares challenge our current understanding of AGN radio variability. Whereas magnetic eruptions offer a viable explanation for the flares, more observations, preferably broadband and including polarisation, will be required to confirm the origin of the flares.}


   \titlerunning{Hour-timescale AGN radio variability}
   \maketitle
\nolinenumbers

%%%%%%%%%%%%%%%%%%%%%%%%%%%%%%%%%%%%%%%%%%%%%%%%%%%%%%%%%%%%%%
\section{Introduction}

Approximately 10\% of active galactic nuclei (AGN) can launch and sustain powerful relativistic jets \citep{2017padovani1}, which are often highly variable yet persistently detectable at radio frequencies. Due to their strong radio emission, jetted AGN have often been identified using the radio loudness parameter, $R${\footnote{Radio loudness parameter, $\emph{R}$, is defined as the ratio between 5~GHz flux density and optical $B$ band flux density. Sources with $\emph{R} >$ 10 are considered radio-loud, $\emph{R} <$ 10 radio-quiet \citep{1989kellermann1}.}. Whereas it is still able to pick out the radio-brightest sources, this parameter fails to represent the true diversity of the jet phenomenon in AGN, spanning from powerful relativistic jets to non-collimated outflows and everything in between \citep{2017padovani1, 2017lahteenmaki1, 2019panessa1, 2022jarvela1}. Moreover, the $R$ parameter often fails to identify even jetted AGN if they are misaligned or low-power. Alternatively, non-jetted AGN whose hosts exhibit strong star formation can be classified as radio-loud \citep{2015caccianiga1, 2022jarvela1, 2022vietri1}. 

Radio-loudness faced another hurdle with the discovery of a small sample of extremely variable radio AGN \citep{2018lahteenmaki1}. Most of these sources are narrow-line Seyfert 1 (NLS1) galaxies \citep{2025crepaldi1}. The NLS1 identification is based on their optical spectrum: the full-width at half maximum (FWHM) of their broad H$\beta$ emission line is $<$2000 km s$^{-1}$, and their [O~III] emission is weak compared to the broad H$\beta$: $S$([O~III])/$S$(H$\beta$)$<$3 \citep{1985osterbrock1, 1989goodrich1}. A large fraction of them also exhibit strong Fe~II emission, confirming the unobstructed view of the central engine. The narrowness of the broad lines can be attributed to the low rotational velocity of the broad-line region (BLR) gas around an intermediate-to-low-mass supermassive black hole (10$^{6}$--10$^{8} M_{\odot}$, \citealp{2011peterson1,2018komossa1}). This interpretation is supported by several lines of evidence, such as reverberation mapping studies \citep{2016wang1,2018du1}, predominantly turbulence-dominated Lorentzian emission line profiles \citep{2011kollatschny1, 2012goad1, 2000sulentic1, 2003marziani1, 2016cracco1, 2020berton1}, the detection of tidal disruption events in NLS1s \citep{2017blanchard1, 2020katebi1, 2021frederick1}, and the prevalence of disk-like host galaxies with pseudo-bulges \citep[e.g.,][]{2011orbandexivry1, 2017jarvela1, 2020olguiniglesias1, 2022varglund1, 2023varglund1}. On average, NLS1s accrete at very high rates (average $L/L_\mathrm{Edd}\sim$ 0.3), often exceeding the Eddington limit \citep{1992boroson1}, as also indicated by their strong Fe~II emission in comparison to the broad H$\beta$ (R4570 or $R_\mathrm{Fe~II}$), known to correlate with the Eddington ratio \citep{2025marziani1}. These properties suggest that NLS1s are fast-growing AGN in an early evolutionary stage \citep{2000mathur1}.

The sample of NLS1s showing extreme variability properties was discovered as part of the Metsähovi Radio Observatory (MRO, Finland) AGN monitoring programme at 37~GHz \citep{2017lahteenmaki1}, and later all of them have also been detected at 15~GHz by the Owens Valley Radio Observatory (OVRO, USA) \citep{2020berton2, 2024jarvela1}. These sources are intriguing because they are often undetectable, but during flares they can reach Jy-level flux densities. NLS1s were found to host relativistic jets in the late 2000's \citep{2006komossa1, 2008yuan1, 2009abdo2}, so the initial hypothesis was that these sources host such jets. One of them was even identified as a gamma-ray emitter and was observed to brighten in X-rays soon after an MRO-detected flare \citep{2023romano1}, supporting the jet hypothesis. However, subsequent medium- and high-resolution radio observations using the Karl G. Jansky Very Large Array (JVLA) and the Very Long Baseline Array (VLBA) did not find any hints of jets \citep{2020berton2, 2021jarvela1, 2024jarvela1}. In VLBA observations at 15~GHz (rms$\sim60~\mu$Jy), none of the sources was detected, ruling out high brightness temperature radio cores seen in other jetted NLS1s, and blazars in general. Multifrequency radio data, including JVLA data from $L$- to $Q$-band, complemented by archival observations down to 144~MHz, revealed that all sources exhibit steep broadband spectra, with several sources undetected at higher frequencies \citep[see Fig. 1 in][]{2024jarvela1}. Only one source, J122844.81+501751.2 (J1228+5017 hereafter), also the focus of this study, showed a flattening of the spectrum at high frequencies, possibly indicating a flat core or a coronal component. The broadband radio spectrum of J1228+5017 is shown in Fig.~\ref{fig:j1228spect}, where the slope appears to flatten above 10~GHz. Despite the absence of detectable jets, \citet{2024jarvela1} report blazar-like variability brightness temperatures ($T_\mathrm{B}\approx$10$^{13}$-10$^{15}$~K) and variability Doppler factors ($\delta\approx$10-30) for one of these sources, which had enough data to perform these estimates. The derived $e$-folding timescales were remarkably short, only of the order of days. The average radio luminosities based on the 37~GHz data are of the order of 10$^{42}$--10$^{43}$~erg~s$^{-1}$.

\cite{2024jarvela1} discuss several possible origins for this extreme behaviour, and after excluding most of them, three alternatives could not be discarded: i) jet or jet base interacting with a star or a BLR cloud; ii) relativistic jets in the presence of extreme free-free absorption; and iii) magnetic reconnection. All of these exhibit distinct multiwavelength characteristics, such as timescales and correlations, particularly in radio and X-rays. Thus, simultaneous radio and X-ray observations could help shed light on the origin of this extreme phenomenon.

To that end, in this study, we analyse simultaneous radio and X-ray observations, complemented by ultraviolet and optical data. In late May 2023, J1228+5017 exhibited a high flux density flare at 37~GHz (780$\pm$120~mJy). Due to this, we triggered the JVLA observations and successfully proposed Target-of-Opportunity observations with the Neil Gehrels \textit{Swift} Observatory \citep[][]{Gehrels2004} to follow up on this flare. OVRO was already monitoring this source. With these data, our goal is to investigate the multiwavelength behaviour of this source; for example, we are interested in whether these bands show any correlated variability, similar to what was previously observed in one of these sources \citep{2023romano1}, or if the higher-energy optical/ultraviolet/X-ray bands show any signs of significant variability. Furthermore, we aim to further constrain the multiband timescales associated with the flaring, since those might help us narrow down the possible explanations for this behaviour. In Sect.~\ref{sec:J1228} we introduce the source, in Sect.~\ref{sec:data} we describe the observations and the data reduction and analysis, in Sect.~\ref{sec:results} we present our results, which are discussed in in Sect.~\ref{sec:discussion}. Finally, we summarise our results and present our conclusions in Sect.~\ref{sec:summary}. Throughout this paper, we adopt a standard $\rm\Lambda CDM$ cosmology, with a Hubble constant $H_0 = 72$~km s$^{-1}$ Mpc$^{-1}$, and $\Omega_\Lambda = 0.73$ \citep{2007spergel1}. For spectral indices, $\alpha$, we adopt the $S_\nu \propto \nu^\alpha$ convention, where $S_\nu$ is the flux density at frequency $\nu$.

\begin{figure}
\begin{center}
\includegraphics[width=\columnwidth]{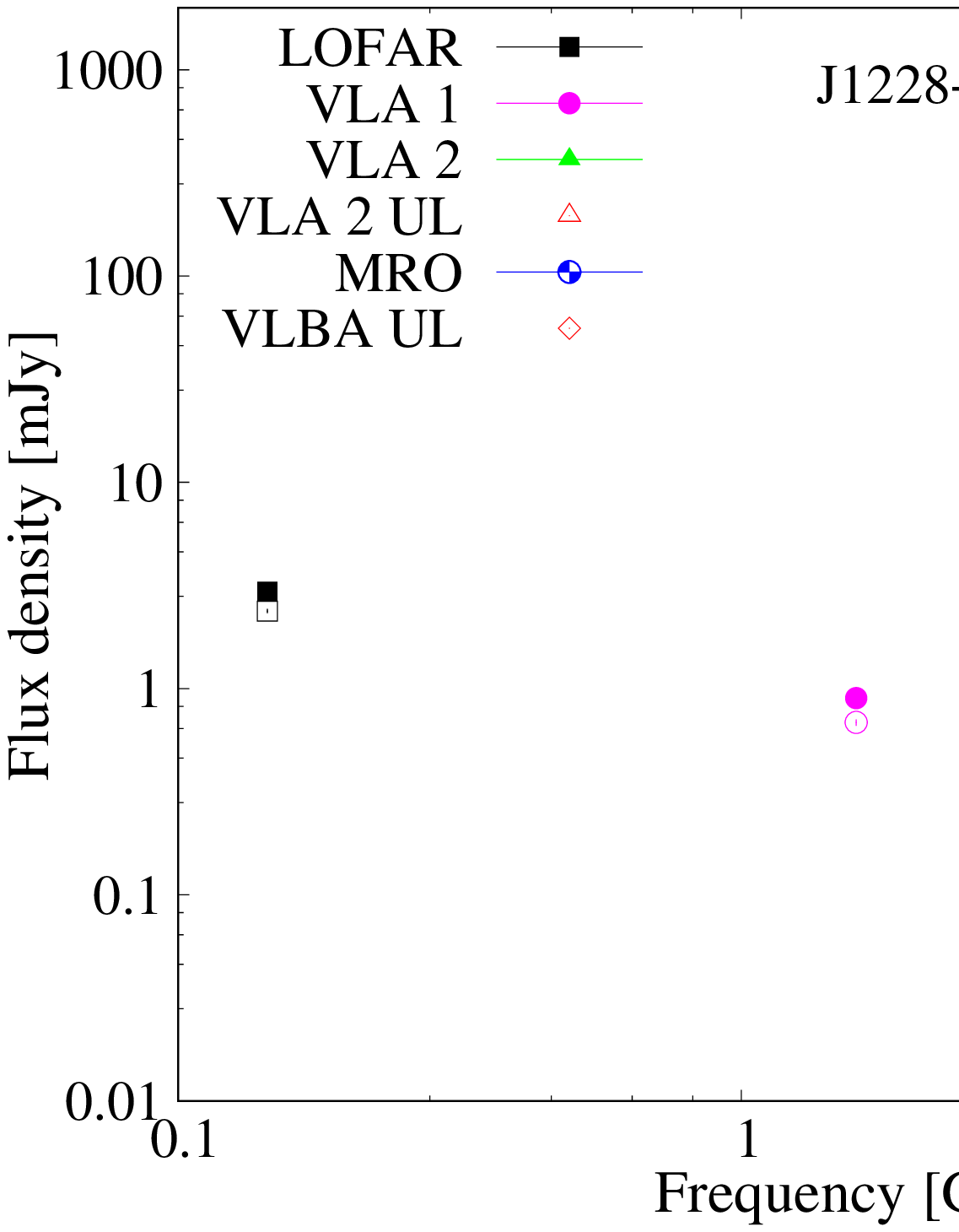}
\caption{Archival broadband radio spectrum of J1228+5017. Symbols and colours are explained in the figure; empty markers denote peak flux densities and filled markers integrated flux densities. VLA 1 and VLA 2 refer to two earlier epochs in 2019 and 2022, and UL = upper limit. Figure from \citet{2024jarvela1}. \label{fig:j1228spect}}
\end{center}
\end{figure}

\section{J1228+5017} \label{sec:J1228}

J1228+5017 (RA 12:28:44.82, Dec +50:17:51.24) is an NLS1 galaxy at a redshift of $z$=0.2627\footnote{Using our adopted cosmology, the spatial scale at this distance is 3.965~kpc~arcsec$^{-1}$.}. It is hosted by a disk-like galaxy \citep{2022varglund1} and was chosen for the MRO monitoring due to its very dense large-scale (Mpc-scale) environment \citep{2017jarvela1, 2018lahteenmaki1}. It was first detected at 37~GHz in 2015, and before the 2023 flare, the last detection was in 2019 \citep{2024jarvela1}. OVRO has also detected it at 15~GHz both before and after the 2023 flare (Crepaldi et al. in prep.). It was also identified as a possible gamma-ray emitter by \citet{2018lahteenmaki1}, who reported a test statistic value of 20, corresponding to a 4.5$\sigma$ tentative detection, yet to be confirmed.

\citet{2025crepaldi1} obtained and modelled a very high-quality (signal-to-noise $\sim$65 in the continuum) optical spectrum of J1228+5017. They estimated its black hole mass to be (9.21 $\pm$ 4.96) $\times$ 10$^6 M_\odot$, and the Eddington ratio to be 0.11 $\pm$ 0.02 or 0.94 $\pm$ 0.19, depending on which method was used to estimate the continuum luminosity. The smaller value results from using the H$\beta_\mathrm{broad}$ luminosity to estimate the monochromatic luminosity at 5100~Å, and the larger value from directly measuring the 5100~Å luminosity. Given its extremely strong Fe~II emission, with an R4570 parameter (strength of Fe~II relative to total H$\beta$) of 2.81, the latter Eddington ratio estimate is more likely to be closer to the true value, as R4570 is known to correlate with the Eddington ratio \citep{2025marziani1}. This interpretation is supported by the strong outflows seen in the [O~III]$\lambda\lambda$4959,5007 lines, often driven by the strong radiation pressure. Modelling these lines requires two wing components, which are blueshifted by $-$163 and $-$371~km~s$^{-1}$, indicating strong outflows \citep[see Fig.~2 of][]{2025crepaldi1}.

There is only one archival X-ray detection of J1228+5017 from ROSAT at 0.0452 counts per second (cps). Thus, we had no prior knowledge of its X-ray properties. However, all in all, J1228+5017 is a stereotypical NLS1 with a low-mass supermassive black hole, high Eddington ratio, strong Fe~II emission, and [O~III] emission lines with strong outflows.

\section{Data} \label{sec:data}

We have simultaneous data from two single-dish telescopes, the JVLA, and the X--ray Telescope \citep[XRT,][]{Burrows2005:XRT} and Ultra-Violet/Optical Telescope \citep[UVOT,][]{Roming2005:UVOT}  onboard \textit{Swift}. The single-dish radio telescopes MRO and OVRO operate at 37 and 15~GHz, respectively. Our JVLA data are in X, K, and Ka bands, centred at 10, 22, and 33~GHz, respectively. XRT data cover the 0.3-10~keV energy band, and our UVOT observations are in V, B, U, W1, M2, and W2 bands. The data included in this work were collected between May 24 and July 13, 2023. 

\subsection{Metsähovi Radio Observatory} 
\label{sec:mrodata}

The measurements included in this study are part of the large ongoing AGN monitoring programme at 22 and 37 GHz with the 13.7-m radio telescope at MRO. The observations are made with a 1 GHz-band dual-beam receiver centred at 36.8~GHz. The beam full-width at half power is 144~arcsec. The observations are on–on observations, alternating the source and the sky in each feed horn. A typical integration time to obtain one flux density data point of a faint source is 1800~s. The sensitivity is limited by sky noise due to the location of the telescope, and it has been experimentally shown that the results do not significantly improve after the maximum integration time of 1800~s. The detection limit of the telescope at 37~GHz is of the order of 200~mJy under optimal conditions. Data points with an S/N $<$ 4 are handled as non-detections. The flux density scale is set by observations of DR 21. Sources NGC 7027, 3C 274, and 3C 84 are used as secondary calibrators. A detailed description of the data reduction and analysis is given in \citet{1998terasranta1}. The error estimate for the flux density includes contributions from the measurement rms and the uncertainty in the absolute calibration. MRO detections for J1228+5017 are shown in Table~\ref{tab:j1228mro} and the MRO light curve, including upper limits, is shown in Fig.~\ref{fig:j1228lc}.

\subsection{Owens Valley Radio Observatory}
\label{sec:ovrodata}

The 15~GHz observations were carried out as part of the general OVRO 40~m radio telescope AGN monitoring programme. This telescope uses off-axis dual-beam optics and a cryogenic receiver with a 15.0~GHz centre frequency and 3~GHz bandwidth. The beam full-width at half power is 157~arcsec. The observations are carried out in an on-on fashion to remove atmospheric and ground contamination. In May 2014, a new pseudo-correlation receiver was installed on the 40~m telescope, and the fast gain variations are corrected using a 180-degree phase switch. Calibration is achieved using a temperature-stable diode noise source to remove receiver gain drifts, and the flux density scale is derived from observations of 3C 286, assuming the \citet{1977baars1} value of 3.44~Jy at 15.0~GHz. The systematic uncertainty of about 5 per cent in the flux density scale is included in the error bars. The upper limits are 4$\sigma$ upper limits based on the measurement rms. Complete details of the reduction and calibration procedure are found in \citet{2011richards1}, and more details specific to the NLS1 observations are given in Appendix A3 \cite{2024jarvela1}. J1228+5017 has been detected by OVRO, but we did not detect it during the period used in this paper. The OVRO upper limits are shown in Fig.~\ref{fig:j1228lc}.

\subsection{Karl G. Jansky Very Large Array} \label{sec:jvladata}

We observed J1228+5017 with the JVLA a total of seven times in three different bands, X, K, and Ka, centred at 10, 22, and 33~GHz, respectively (Project VLA/23A-043, PI Järvelä). The dates and times of the JVLA observations are given in Table~\ref{tab:j1029interf}. The array configuration changed over our observing period: for the first three epochs, the array was in BnA, the next three in BnA$\rightarrow$A, and the last observation in A configuration. The resolutions of these configurations are comparable, and the source was point-like in most observations; thus, we have no reason to assume that configuration changes would have significantly affected the measured flux densities. The total bandwidth was 4~GHz in X, and 8~GHz in K and Ka bands, each band divided into 128~MHz subbands, consisting of 64 channels of 2~MHz. The NLS1 3C 286 \citep{2017berton1} was used as the bandpass and flux density calibrator for each source, and each source had an individual nearby bright source used as the complex gain calibrator. The pointing offset calibration was done either at 3C 286 or the current complex gain calibrator. The expected thermal noise levels were 13, 18, and 22~$\mu$Jy in X, K, and Ka bands, respectively. In most cases, we reached even lower rms levels (see Table~\ref{tab:j1029interf}).

We used the Science Ready Data Products (SRDP) provided by the NRAO. The data were calibrated using the VLA Imaging Pipeline 2022.2.0.64. After this, the data were checked manually, and any remaining bad data were flagged, producing the SRDP measurement set for each source. We also performed a manual check of the data, but no additional flagging was required. In the subsequent steps, we used the Common Astronomy Software Applications (CASA) version 6.4.1.12. We averaged the J1228+5017 data in time (\texttt{timebin} = 10~s) and in frequency (\texttt{width} = 64, averaging 64 channels to form one output channel per spectral window). % Did I check the large area? Probably?

The radio maps of our sources were produced using the \texttt{tclean} algorithm with interactive cleaning. The cell size was chosen to ensure the synthesised beam is properly sampled; we used cell sizes of 0.025, 0.010, and 0.007~arcsec in the X, K, and Ka bands, respectively. The image size in the X band was 512$\times$512 pixels, whereas we used 1024$\times$1024 in the K and Ka bands. We used Briggs weighting, with \texttt{robust} = 1.8, in all cases. We used the \texttt{mtmfs} deconvolver with \texttt{nterms} = 2 and \texttt{scales} = 0 in case J1228+5017 would be bright enough to produce spatially resolved in-band spectral index maps, which turned out not to be the case. We applied the wide-band primary beam correction separately using \texttt{widebandpbcor}.

We fitted the source in each map using a 2D Gaussian to obtain the central coordinates, the peak flux density, and its error. In the case of extended emission (only two Ka band observations), we measured the emission within the 3$\sigma$ contour and estimated its error by multiplying the rms by the square root of the emitting region area, expressed in beams. The rms for each map was measured in an empty region of sky far from the central source. Details of the measurements are given in Table~\ref{tab:j1029interf}, the radio spectra for the seven epochs are shown in Fig.~\ref{fig:j1228jvlaspectra}, and the JVLA light curves are shown in Fig.~\ref{fig:j1228lc}. Since the JVLA flux density scale calibration accuracy is of order 5\% in X band, and 10-15\% in K and Ka bands, we took this into account in the error bars shown in Fig.~\ref{fig:j1228jvlaspectra} by adding in quadrature. The errors reported in Table~\ref{tab:j1029interf} as well as the errors shown in Fig.~\ref{fig:j1228lc} do not include this additional source of error.

\subsection{Neil Gehrels \textit{Swift} Observatory} %

The \textit{Swift} data were collected through target of opportunity observations (Target ID 16062, PI: P.\ Romano, 13 observations) from 2023-05-30 to 2023-07-04 with XRT \citep{Burrows2005:XRT} and the UVOT \citep[][with the ``U+B+V+all U'' filter setup]{Roming2005:UVOT}, for a total exposure of 34.7~ks. 

The \textit{Swift}/XRT data were reduced and analysed by using standard procedures within {\sc FTOOLS}\footnote{\href{https://heasarc.gsfc.nasa.gov/ftools/ftools_menu.html}{https://heasarc.gsfc.nasa.gov/ftools/ftools\_menu.html.}} (v6.36), and the matching calibration database CALDB\footnote{\href{https://heasarc.gsfc.nasa.gov/docs/heasarc/caldb/caldb_intro.html}{https://heasarc.gsfc.nasa.gov/docs/heasarc/caldb/caldb\_intro.html.}}.  In particular, the XRT data were processed and filtered with the task {\sc xrtpipeline} (v0.13.7). The XRT light curve was generated in the 0.3--10~keV energy range with one point per observation. We used {\sc sosta} within {\sc XIMAGE} with a background measured in an annular region with an inner radius of 60 pixels and an outer radius of 110 pixels centred at the source position. The appropriate exposure maps were used to correct for PSF losses and vignetting. We also extracted a total spectrum by adding all XRT observations (34.7~ks). For the spectroscopic analysis, source events were extracted from a circular region with a 15-pixel radius; ancillary response files were generated with the task {\sc xrtmkarf} to account for different extraction regions, vignetting, and PSF corrections; the spectral redistribution matrices in CALDB were used. We fit the total spectrum with {\sc XSPEC} \citep[v.\ 12.15.1,][]{Arnaud1996:xspec} with a {\sc tBabs*zTBabs*powerlaw}, that is a simple power law model corrected for an equivalent hydrogen column corresponding to the Galactic value (fixed $N_{\rm H}^{\rm Gal}=1.56\times10^{20}$~cm$^{-2}$, \citealt[][]{NH2016:HI4PI}) and with the {\sc tbabs} \citep[][]{Wilms2000} model and a further absorption column at fixed redshift $z=0.262$. We obtain a photon index $\Gamma= 2.33^{+0.20}_{-0.18}$ and an observed (unabsorbed) 0.3-10~keV flux of $(6.6\pm{0.6}) \times10^{-13}$ ($7.5^{+0.9}_{-0.7} \times10^{-13}$) erg~cm$^{-2}$~s$^{-1}$  (statistics/dof= 156.36/234). 

The analysis of the UVOT data, collected in all filters (optical $v$, $b$, $u$ and UV $w1$, $m2$, $w2$), was performed with the tasks {\sc uvotmaghist},   {\sc uvotimsum}, and {\sc uvotsource} within {\sc FTOOLS} (v6.34 which offer a check for frames affected by small scale sensitivity, SSS, issues\footnote{\href{https://swift.gsfc.nasa.gov/analysis/uvot_digest/sss_check.html}{https://swift.gsfc.nasa.gov/analysis/uvot\_digest/sss\_check.html}.}). We used {\sc uvotmaghist} to identify the frames of each UVOT image affected by SSS issues and to generate a light curve in each filter at the frame level (the best time resolution available), {\sc uvotimsum} to sum the sky images for {\sc uvotsource} to calculate the magnitude of the source through aperture photometry within a circular region centred on the best source position by applying the required corrections related to the specific detector characteristics. We adopted circular regions with radii of 5~\arcsec{} and 20~\arcsec{} for the source and background, respectively. 

The \textit{Swift}/XRT observing epochs and exposure times are shown in Table~\ref{tab:swift_xrt_log}, and the light curves for XRT and UVOT can be found in Fig.~\ref{fig:j1228lc}.

\begin{figure*}
\begin{center}
\includegraphics[trim={0 0 0 3.0cm}, clip, width=15cm]{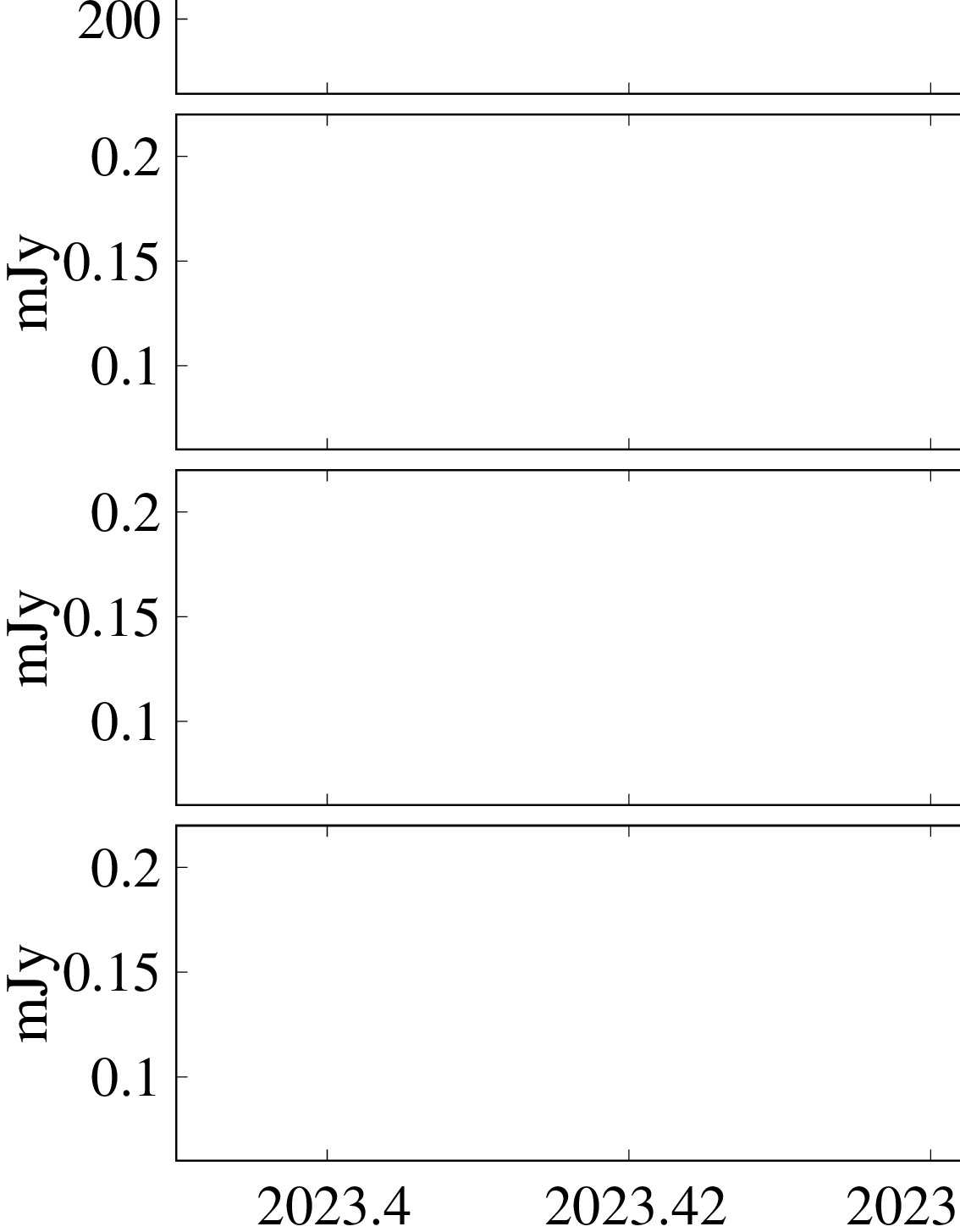}
\caption{Multiwavelength light curves of J1228+5017. Symbols and colours are explained in the figure. \label{fig:j1228lc}}
\end{center}
\end{figure*}

\begin{figure}
\begin{center}
\includegraphics[width=\columnwidth]{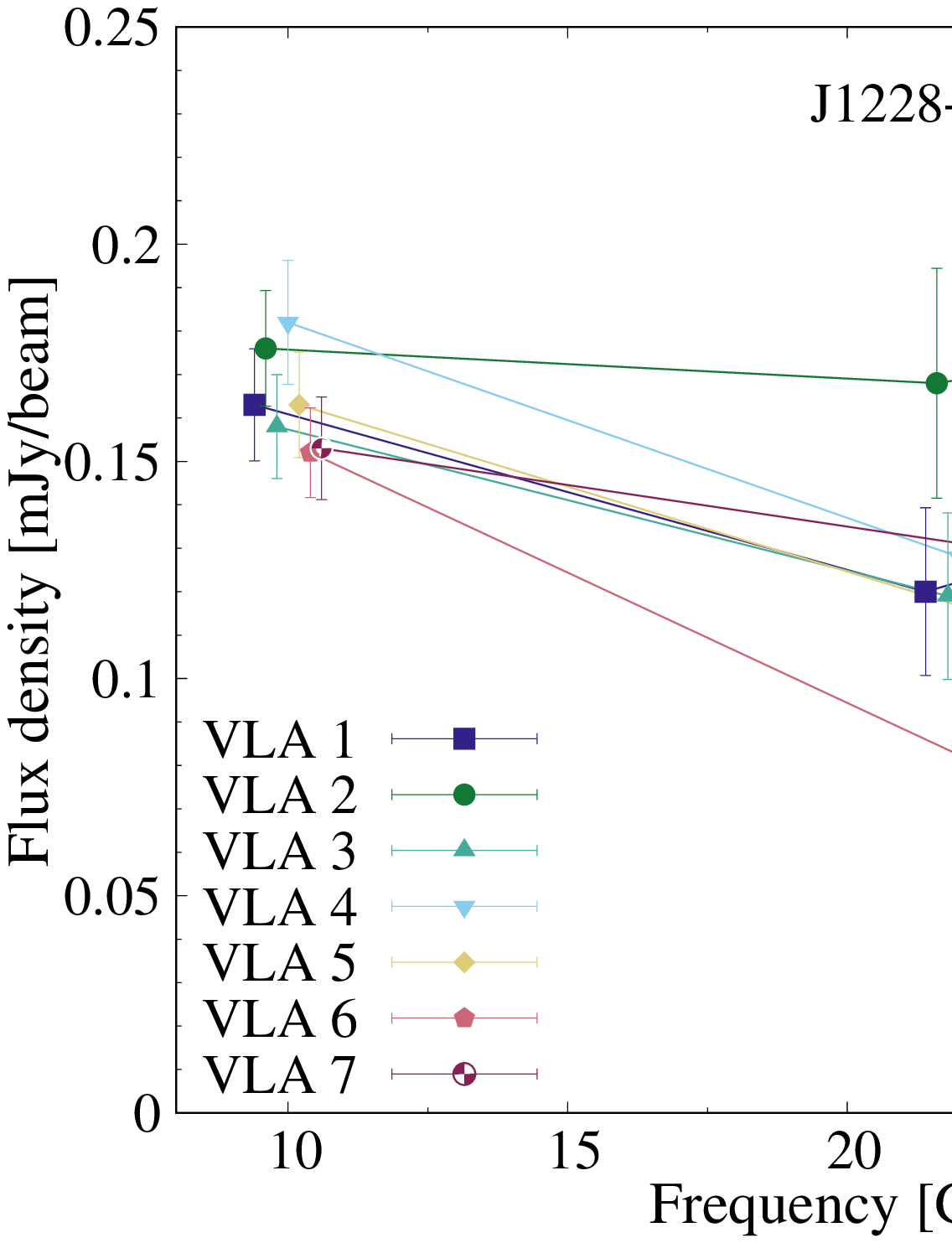}
\caption{JVLA radio spectrum of J1228+5017. Symbols and colours are explained in the figure. The JVLA flux density scale calibration error has been accounted for in the error bars. The central frequencies have been changed slightly for readability. The epoch numbers correspond to those shown in Table~\ref{tab:j1029interf}. \label{fig:j1228jvlaspectra}}
\end{center}
\end{figure}

\section{Results} \label{sec:results}

\subsection{Note on measured variability}

Before presenting the results, we briefly discuss the variability measures and statistics used. We performed the weighted $\chi^2_\mathrm{r}$ test, as well as calculated the $C$ and $F$ parameters for all light curves \citep{2020singh1}. These tests indicate that only the XRT and W2 light curves are variable. However, the number of data points in the light curves ranges from seven to 13, leading to low degrees of freedom. This directly determines the $\chi^2$ distribution and adversely impacts the power and validity of these tests. Moreover, 7-13 observations cannot properly represent the underlying flux density distribution, and generally it is not advised to use statistical tests based on the normal distribution of the data with fewer than 30 data points. With that being said, based on these tests, we cannot rule out the possibility that all the light curves, except for XRT and W2, are constant. However, several light curves do \textit{appear} variable, as we will discuss in the following sections. 

\subsection{Single-dish observations}

All the follow-up observations were triggered by an MRO detection on May 28, 2023. The flux density of this detection was 780 $\pm$ 120~mJy. During the observing period shown in Fig.~\ref{fig:j1228lc}, J1228+5017 was detected a total of nine times, with $S_\nu,\mathrm{max}$ = 870 $\pm$ 170~mJy, $S_\nu,\mathrm{min}$ = 440 $\pm$ 80~mJy and $S_\nu,\mathrm{ave}$ = 600~mJy. During this period, J1228+5017 was observed 61 times, yielding a detection rate of 15\%. Upper limits were derived for the 52 non-detections. The MRO light curve shows repeating flaring behaviour at 37~GHz; on average, the upper limits are lower than the detections, suggesting the source would have been detected had it maintained its high flux density. However, based on the 37~GHz observations, we cannot say much about the flux density levels between the detections; in principle, the source could have been just below the detection limit.

Interestingly, OVRO did not detect the source during this period. The 15~GHz observations yielded only five upper limits, on the order of a few tens of mJy. On June 15, an OVRO non-detection was followed by an MRO detection only 1.2~hours later, indicating a very short timescale, assuming that this phenomenon is broadband. The OVRO beam size is close to that of MRO, so OVRO upper limits can be used to estimate the low flux density baseline for the MRO flux densities by assuming some reasonable spectral index between the two frequencies. These will be discussed more in Sect.\ref{sec:timescale}.

\begin{table}
\caption{MRO 37~GHz single-dish detections for J1228+5017. Columns: (1) date of the observation; (2) flux density and its error.}
\label{tab:j1228mro}
\centering
\begin{tabular}{lcc}
\hline\hline
date        & date and time         & $S_{\mathrm{int}}$  \\
(dec. year) & (yyyy-mm-dd hh:mm:ss) & (mJy)    \\  \hline
2023.405411	& 2023-05-28 23:23:53   & 780 $\pm$ 120 \\
2023.427261	& 2023-06-05 22:48:19   & 440 $\pm$ 80 \\
2023.432697	& 2023-06-07 22:25:21   & 500 $\pm$ 110 \\
2023.442703	& 2023-06-11 14:04:43   & 870 $\pm$ 170 \\
2023.443627	& 2023-06-11 22:10:18   & 610 $\pm$ 120 \\
2023.450887	& 2023-06-14 13:46:12   & 630 $\pm$ 150 \\
2023.454555	& 2023-06-15 21:53:54   & 600 $\pm$ 130 \\
2023.476434	& 2023-06-23 21:33:45   & 490 $\pm$ 90 \\
2023.506495	& 2023-07-04 20:54:00   & 480 $\pm$ 90 \\ \hline
\end{tabular}
\end{table}

\subsection{Interferometric observations}

Our arcsec-resolution radio data consist of seven epochs of JVLA observations in the X, K, and Ka bands from June 14 to July 1, 2023. J1228+5017 was detected at $> 6\sigma$ level in all bands at all epochs (see Table~\ref{tab:j1029interf}. It was otherwise consistent with a point-like morphology except for the last two Ka band observations, where it was marginally resolved. However, within the errors, its integrated flux density remains consistent with the corresponding peak flux density, indicating that the amount of any resolved emission is minimal. JVLA also changed its configuration from BnA to A during these observations, with the last observations having the highest resolution. There are no trends of decreasing flux density moving from BnA to A configuration, ruling out resolved-out emission having a significant impact on the observed variability.

All bands seem to show variability over the observing period, even if, based on the variability test, a constant flux density cannot be ruled out. The X band shows the least change in flux density, but even in this case, the minimum and maximum flux densities differ, even when accounting for the flux density scale accuracy, as can be seen in Fig.~\ref{fig:j1228spect}. The variability is more pronounced in K and Ka bands, with the maximum flux density being approximately double the minimum flux density. As mentioned earlier, neither the array configuration nor the clean beam size appears to correlate significantly with flux density.

The average peak flux densities in X, K, and Ka bands are 164 $\pm$ 4, 123 $\pm$ 10, and 130 $\pm$ 13~$\mu$Jy~beam$^{-1}$, respectively. We estimate the error as the standard error of the mean. Compared to the peak flux densities reported by \cite{2024jarvela1}, the X and Ka band values are statistically higher; they report X, K, and Ka band peak flux densities of 128 $\pm$ 5, 120 $\pm$ 7, and 102 $\pm$ 11~$\mu$Jy~beam$^{-1}$.

As a result of possible variability in the K and Ka bands, the spectral shape in the 10-33~GHz range also appears to change, showing flat, concave, and steep shapes. Still, even at its steepest at epoch 5, the spectral index is $\alpha$ = $-$0.44, and thus the spectrum is formally flat\footnote{Based on the spectral index, a spectrum is steep when $\alpha < -0.5$, inverted when $\alpha > 0.5$, and flat between these values.}. This is consistent with the earlier broadband spectrum shown in Fig.~\ref{fig:j1228spect}, which indicates that the spectrum flattens above 10~GHz.

There are two instances where we have detection at 37~GHz from MRO and in Ka band from JVLA within 12~hours of each other. These consecutive observations will be discussed in Sect.~\ref{sec:timescale}.

\begin{table*}
\caption{Interferometric data for J1228 in X, K and Ka bands. Columns: (1) date in decimal years; (2) date and the observation end time; (3) peak flux density and its error (from the 2D Gaussian fit), not including the flux density calibration uncertainty; (4) integrated flux density and its error; (5) rms level of the observation; (6) clean beam size; (7) beam position angle; (8) array configuration.}
\label{tab:j1029interf}
\centering
  \small % < ------------------------------------------ 
\begin{tabular}{lccccccc}
\hline\hline
              \multicolumn{8}{c}{X}                                                                        \\
date        & date and time         & $S_{\mathrm{peak}}$ & $S_{\mathrm{int}}$ & rms                   & beam size          & beam PA & array   \\
(dec. year) & (yyyy-mm-dd hh:mm:ss) & ($\mu$Jy beam$^{-1}$)           & ($\mu$Jy)          & ($\mu$Jy beam$^{-1}$) & (arcsec)           & (deg)   & configuration   \\  \hline
2023.449861 (VLA 1) & 2023-06-14 04:47:11   & 163 $\pm$ 10        &                    & 12                    & 0.64 $\times$ 0.49 & 67.8    & BnA \\
2023.449970 (VLA 2) & 2023-06-14 05:44:08   & 176 $\pm$ 10        &                    & 12                    & 0.74 $\times$ 0.43 & 60.0    & BnA \\
2023.452638 (VLA 3) & 2023-06-15 05:06:30   & 158 $\pm$ 9         &                    & 11                    & 0.66 $\times$ 0.48 & 66.0    & BnA \\
2023.471926 (VLA 4) & 2023-06-22 06:04:25   & 182 $\pm$ 11        &                    & 13                    & 0.94 $\times$ 0.26 & 52.2    & BnA$\rightarrow$A\\  
2023.488120 (VLA 5) & 2023-06-28 03:56:06   & 163 $\pm$ 9         &                    & 11                    & 0.26 $\times$ 0.22 & $-64.3$ & BnA$\rightarrow$A \\
2023.488229 (VLA 6) & 2023-06-28 04:53:02   & 152 $\pm$ 7         &                    & 13                    & 0.28 $\times$ 0.21 & $-76.3$ & BnA$\rightarrow$A \\      
2023.496394 (VLA 7) & 2023-07-01 04:24:50   & 153 $\pm$ 9         &                    & 11                    & 0.26 $\times$ 0.22 & $-71.8$ & A \\     \hline 
              \multicolumn{8}{c}{K}                                                                        \\
date        & date and time         & $S_{\mathrm{peak}}$ & $S_{\mathrm{int}}$ & rms                   & beam size          & beam PA & array   \\
(dec. year) & (yyyy-mm-dd hh:mm:ss) & ($\mu$Jy beam$^{-1}$)           & ($\mu$Jy)          & ($\mu$Jy beam$^{-1}$) & (arcsec)           & (deg)   & configuration   \\  \hline
2023.449861 (VLA 1) & 2023-06-14 04:47:11   & 120 $\pm$ 7         &                    & 10                    & 0.29 $\times$ 0.22 & 71.5    & BnA    \\
2023.449970 (VLA 2) & 2023-06-14 05:44:08   & 168 $\pm$ 8         &                    & 11                    & 0.30 $\times$ 0.21 & 71.1    & BnA    \\
2023.452638 (VLA 3) & 2023-06-15 05:06:30   & 119 $\pm$ 7         &                    & 10                    & 0.30 $\times$ 0.24 & 63.1    & BnA    \\
2023.471926 (VLA 4) & 2023-06-22 06:04:25   & 128 $\pm$ 12        &                    & 15                    & 0.38 $\times$ 0.12 & 54.0    & BnA$\rightarrow$A   \\  
2023.488120 (VLA 5) & 2023-06-28 03:56:06   & 116 $\pm$ 9         &                    & 11                    & 0.12 $\times$ 0.09 & $-64.8$ & BnA$\rightarrow$A    \\
2023.488229 (VLA 6) & 2023-06-28 04:53:02   & 80  $\pm$ 11        &                    & 13                    & 0.13 $\times$ 0.10 & $-77.7$ & BnA$\rightarrow$A    \\      
2023.496394 (VLA 7) & 2023-07-01 04:24:50   & 130 $\pm$ 10        &                    & 10                    & 0.12 $\times$ 0.09 & $-68.4$ & A    \\     \hline
              \multicolumn{8}{c}{Ka}                                                                        \\
date        & date and time         & $S_{\mathrm{peak}}$ & $S_{\mathrm{int}}$ & rms                   & beam size          & beam PA  &  array    \\
(dec. year) & (yyyy-mm-dd hh:mm:ss) & ($\mu$Jy beam$^{-1}$)           & ($\mu$Jy)          & ($\mu$Jy beam$^{-1}$) & (arcsec)           & (deg)    & configuration  \\ \hline
2023.449861 (VLA 1) & 2023-06-14 04:47:11   & 161 $\pm$ 11        &                    & 14                    & 0.19 $\times$ 0.16 & 79.1     & BnA  \\
2023.449970 (VLA 2) & 2023-06-14 05:44:08   & 190 $\pm$ 13        &                    & 14                    & 0.20 $\times$ 0.14 & 70.6     & BnA  \\
2023.452638 (VLA 3) & 2023-06-15 05:06:30   & 110 $\pm$ 11        &                    & 13                    & 0.20 $\times$ 0.17 & 61.7     & BnA  \\
2023.471926 (VLA 4) & 2023-06-22 06:04:25   & 135 $\pm$ 13        &                    & 16                    & 0.26 $\times$ 0.08 & 53.8     & BnA$\rightarrow$A   \\  
2023.488120 (VLA 5) & 2023-06-28 03:56:06   & 96  $\pm$ 11        &                    & 15                    & 0.08 $\times$ 0.06 & $-61.7$  & BnA$\rightarrow$A   \\
2023.488229 (VLA 6) & 2023-06-28 04:53:02   & 121 $\pm$ 11        & 130 $\pm$ 20       & 15                    & 0.08 $\times$ 0.06 & $-77.9$  & BnA$\rightarrow$A   \\      
2023.496394 (VLA 7) & 2023-07-01 04:24:50   & 100 $\pm$ 11        & 120 $\pm$ 18       & 13                    & 0.09 $\times$ 0.06 & $-67.2$  & A  \\     \hline 
\end{tabular}
\end{table*}

\subsection{Optical, ultraviolet, and X-ray observations}

We have 13 epochs of \textit{Swift} data obtained with XRT and UVOT. The XRT energy range is 0.3--10~keV, and the UVOT bands used are V, B, U, W1, M2, and W2, with central wavelengths of 5468, 4392, 3465, 2600, 2246, and 1928~Å, respectively. The light curves are shown in Fig.~\ref{fig:j1228lc}.

The optical and UV light curves seem to show variability by a factor of 1.5-2, but we identify no obvious flare-like events. V and B bands seem to be the most variable, which might be due to the fact that the Balmer discontinuity falls in the B band, and the H$\gamma$ and approximately half of the blueward Fe~II bump fall in the V band. However, based on the archival optical spectra of J1228+5017, its spectrum has stayed remarkably similar over the past 20 years. The only statistically variable optical/UV light curve is the W2 band; this is likely due to its very small errors, as variability tests are highly sensitive to the size of the measurement errors.

The X-ray light curve is slightly more variable, also according to the $\chi^2_\mathrm{r}$ variance test, with variability by a factor of $\sim$3. Similar to the V and B band light curves, the X-rays seem to be more elevated during the first half of the observing period, but no flare-like events with unusually high amplitude can be seen. We observe no clear, similar trends when comparing the optical, UV, or X-ray light curves to the radio light curves.

\subsection{Variability timescales} \label{sec:timescale}

We can use consecutive detections, or a nearby detection and an upper limit, to estimate variability properties, such as timescales, variability brightness temperatures, and variability Doppler factors of the flares \citep{1999valtaoja1, 2009hovatta1}. In the following, we assume that the flux density differences between bands and/or instruments trace the evolution of the same variable component. We will consider the following cases: i) MRO 37~GHz detection and JVLA Ka band detection within 24~hours of each other; and ii) MRO detection and OVRO non-detection within 24~hours of each other. We will also consider two consecutive JVLA or MRO detections, but in those cases, the flux densities are similar when accounting for the errors. Before performing any calculations, we make some remarks on these cases and explain the assumptions we will use.

When using MRO and JVLA data, the resolutions differ significantly: 144~arcsec for MRO and much less than 1~arcsec for the JVLA in the Ka band (33~GHz). The largest angular scale detectable with the JVLA in A configuration in Ka band is 1.6~arcsec, which translates to 6.3~kpc at the redshift of our source. We do not know how much of the emission is resolved out with the JVLA, which is why, in these cases, we will use the closest OVRO upper limit and a spectral index of $\alpha$ = 0 to estimate the baseline flux density at 37~GHz. This approach is based on the similar beam sizes of MRO and OVRO --- they should probe approximately the same radio-emitting regions, with the flux density modulated by the spectral index --- and on the fact that, due to its spatial extent, the resolved-out emission cannot vary on short timescales. We adopt a spectral index of 0, which is close to the spectral slope of free-free emission (i.e. $-$0.1). As far as we know, there are no emission mechanisms that produce inverted spectra at kpc-scales. For JVLA flux density errors, we will use values that account for the flux density scale uncertainty.

Since MRO and OVRO are observing at different frequencies, when comparing an MRO detection and an OVRO upper limit, we will extrapolate the OVRO upper limit to 37~GHz using an extreme spectral index of $\alpha$ = 4. We adopt this value to represent the worst-case scenario, for example, extreme synchrotron self-absorption coupled with external free-free absorption. In this case, we assume the flares are broadband, which may not be true. We also want to remind the reader that upper limits are not detections and should not be treated as such; the results using an upper limit and extrapolation of data will be less accurate and reliable than the ones derived with detections.

With these caveats and assumptions in mind, we can estimate the timescales, variability brightness temperatures, and variability Doppler factors of the flares. For simplicity, we used the same equation for both rising and decaying flares:
\begin{equation}
\Delta S (t) = \Delta S_{\textrm{max}} \textrm{e}^{(\vert{}t-t_{\textrm{max}}\vert{}) / \tau} \quad [\mathrm{Jy}], \label{eq:tau}
\end{equation}
where $\Delta S_{\textrm{max}}$ is the maximum amplitude of the flare in Jy, after subtracting the baseline flux density level, $S_{\textrm{b}}$, $t_{\textrm{max}}$ is the epoch of the peak of the flare, and $\tau$ is the rise or decay time of the flare expressed in days ($e$-folding timescale). In our case, we only have two measurements per flare, so we can simplify $\vert{}t-t_\mathrm{max}\vert{}$ to $\Delta t$, which is the time between the middle points of the observations.

The variability brightness temperature, $T_{\textrm{b, var}}$, in the source proper frame was estimated with:
\begin{equation}
T_{\textrm{b, var}} = 1.548 \times 10^{-32} \frac{\Delta S_{\textrm{max}} d_L^2}{\nu^2 \tau^2 (1+z)} \quad [\mathrm{K}],
\end{equation}
where $\nu$ is the observed frequency in GHz, $d_{L}$ is the luminosity distance in metres, and $\Delta S_{\textrm{max}}$ and $\tau$ are defined in Eq.~\ref{eq:tau}. The numerical factor corresponds to using $H_0 = 72$~km s$^{-1}$ Mpc$^{-1}$, and $\Omega_\Lambda = 0.73$, and assuming that the source is a homogeneous sphere. Estimating the brightness temperature from flux density variability relies on a causality argument, so these values are lower limits. We calculated the variability brightness temperatures for all flares with $\tau$ values. We note that the brightness temperatures derived from variability are systematically larger by a factor of $\delta^3$, where $\delta$ is the Doppler factor, than those obtained directly from high-resolution measurements due to the different dependence on the Doppler factor. 

Finally, we can use the variability brightness temperature to estimate the variability Doppler factor, assuming we know the intrinsic brightness temperature, $T_{\textrm{b, int}}$:
\begin{equation}
\delta_{\textrm{var}} = \left( \frac{T_{\textrm{b, var}}}{T_{\textrm{b, int}}} \right) ^{1/3} .
\end{equation}

For the intrinsic brightness temperature, we use 5 $\times$ 10$^{10}$~K \citep{1994readhead1, 1999lahteenmaki1, 2021homan1}, which assumes equipartition between the energy densities of the magnetic field and the radiating particles. We do not know whether J1228+5017 is in equipartition and therefore cannot assess the accuracy of the Doppler factor estimates.

The $e$-folding timescales, variability brightness temperatures and variability Doppler factors for three flares are shown in Table~\ref{tab:flaretau}. In two cases, J1228+5017 was detected in Ka band (33~GHz) and shortly after at 37~GHz. In one case, a 37~GHz detection was followed by a 15~GHz non-detection only 1.2~h later. The $e$-folding timescales in all cases are on the order of hours, even when accounting for errors. The derived variability brightness temperatures are of the order of $10^{17}-10^{18}$~K. These values far exceed the Compton catastrophe threshold of 10$^{12}$~K, but can be partly explained by extreme Doppler factors, due to the dependence of the $\delta^3$ term. Resulting from the high variability brightness temperatures, the derived variability Doppler factors are of the order of a few hundred.

Using a simple light-travel-time argument to estimate the size of the radio-emitting region, its size needs to be $r < c \tau \delta / (1 + z)$. For the shortest $e$-folding timescale, this gives $r <$ 0.03~mpc $\times~\delta$. If the Doppler factor exceeds one hundred, the emitting region would be on the order of a few mpc, or 3-4 light-days; comparable to the size of the accretion disk. Adopting the \citet{2025crepaldi1} value for the black hole mass, 9.21 $\times 10^6 M_\odot$, the gravitational radius, $r_\mathrm{g}$, of J1228+5017 is 4.4 $\times$ 10$^{-4}$~mpc. It should be remembered that the derived timescales are upper limits, and the actual variability might occur faster, indicating smaller emitting regions.

\begin{table*}
\tabcolsep 5pt
\caption{Flare properties from the MRO data. Columns: (1) facilities used; (2) date of the observations; (3) time difference between the midpoints of the observations; (4, 5) flux density at the beginning and at the end of the flare; (6) timescale; (7) variability brightness temperature; (8) variability Doppler factor.}
\label{tab:flaretau}
\centering
\renewcommand{\arraystretch}{1.5}
\begin{tabular}{lccccccc}
\hline\hline
facilities   & date         & $\Delta t$ & $S_{\textrm{begin}}$      & $S_{\textrm{end}}$        & $\tau$                                 & $T_{\textrm{b, var}}$             & $\delta_{\textrm{var}}$ \\ 
             & (yyyy-mm-dd) & (hours)    & (mJy)                     &  (mJy)                    & (hours)                                 & (10$^{18}$~K)                    & \\ \hline
VLA Ka -- MRO & 2023-06-14   & 8.9672     & 0.19 $\pm$ 0.03$^{a}$     & 618.00$^{b}$ $\pm$ 150.00 & 1.1088 $\substack{+0.0615 \\ -0.0509}$  & 4.20 $\substack{+1.53 \\ -1.35}$ & 438 $\substack{+48 \\ -53}$  \\
VLA Ka -- MRO & 2023-06-15   & 17.72278   & 0.11 $\pm$ 0.02$^{a}$     & 588.00$^{b}$ $\pm$ 130.00 & 2.0646 $\substack{+0.1054 \\ -0.0920}$  & 1.15 $\substack{+0.40 \\ -0.34}$ & 284 $\substack{+30 \\ -31}$ \\
MRO -- OVRO   & 2023-06-15   & 1.1919     & 588.00$^{b}$ $\pm$ 130.00 & 424.00$^{c}$              & 3.6450 $\substack{+11.8070 \\ -1.3822}$ & 0.37 $\substack{+0.80\\ -0.35}$  & 195 $\substack{+91 \\ -121}$   \\ \hline 
\end{tabular} \\
{\footnotesize $^{a}$ Error includes the flux density scale uncertainty; $^{b}$ Base flux density, calculated from the nearest OVRO upper limit, has been subtracted; $^{c}$ upper limit at 37~GHz, estimated from the 15~GHz upper limit of 11.45~mJy using $\alpha$ = 4.}
\end{table*}

\section{Discussion} \label{sec:discussion}

A thorough discussion of alternative explanations, intrinsic and external to the source, is given by \citet{2024jarvela1} and will not be repeated here. Instead, we focus on interpreting the multiwavelength monitoring of J1228+5017 and the extremely short timescales, coupled with high-amplitude variability, $T_\mathrm{B, var}$ and $\delta_\mathrm{var}$.

\subsection{Silence of the JVLA}

The JVLA observations pose two interesting questions. First, why are these flares not detected by the JVLA, and second, what is the origin of the high-frequency flat-spectrum radio emission seen by the JVLA?

The JVLA observed J1228+5017 seven times over approximately two weeks. It coincides with a period when MRO observed the source 28 times and detected it three times, giving a detection fraction of $\sim$10\% (see \citealp{2024jarvela1} for a thorough discussion about the reliability of the MRO detections). With a 10\% duty cycle, it is not an unreasonable assumption that the seven JVLA observations just missed the flares. However, given MRO's moderate sensitivity, we would expect a higher duty cycle when observed with the higher-sensitivity JVLA, since it should also detect rising or decaying flares that remain below MRO's detection threshold. Since this is not the case, it may indicate even shorter timescales or flare profiles not well represented by exponential rise and decay. The nature of the flares is discussed in more detail in Sect.~\ref{sec:natureofflares}.

Given the low flux densities between 10 and 33~GHz, it is worth exploring whether star formation, rather than an AGN, could be responsible for the radio emission. At these frequencies, arcsec-scale AGN radio spectra are often dominated by a steep ($\alpha \approx -0.7$) optically thin synchrotron component \citep{2019panessa1}, whereas in the presence of star formation, toward high frequencies, the flat free-free emission ($\alpha \approx -0.1$) starts to flatten the steep supernova remnant synchrotron spectrum \citep[$\alpha \approx -0.8$][]{2018klein1}. This spectrum often also has a break around 1-12~GHz, further decreasing its contribution at higher frequencies \citep{2018klein1}. At 33~GHz, the thermal fraction is already around 80\% \citep{2025dignan1}, indicating that the radio spectrum should be flat. This alone does not prove that star formation is present; thus, we will utilise additional star formation proxies to estimate its contribution. First, we can estimate the star formation rate (SFR) required to produce the 33~GHz radio emission following Eq. 10 by \citet{2012murphy1}. We assume $T_\mathrm{e}$ = 10$^4$~K and a non-thermal spectral index, $\alpha_\mathrm{NH}$ = $-$0.7. We calculate the luminosity, $L_\nu$, from the average flux density of the last three epochs (105.7 $\pm$ 19.1)~$\mu$Jy, since these observations have the smallest clean beam, thus providing the best constraints on the size of the emitting region. Based on the resulting luminosity, $L_\mathrm{33~GHz}$ = (1.70 $\pm$ 0.3) $\times 10^{29}$~erg~s$^{-1}$~Hz$^{-1}$, the SFR would be $\sim$60$\pm$10~$M_\odot$~yr$^{-1}$ within the innermost (0.36 $\times$ 0.24)~kpc$^2$. This would yield an SFR surface density of $\sim$694~$M_\odot~$y$^{-1}$~kpc$^{-2}$, which is unreasonably high compared to the average values in the local Universe, which are around 0.5~$M_\odot~$y$^{-1}$~kpc$^{-2}$ \citep{2022winkel1}, and thus it is unlikely that star formation has a significant contribution to the radio emission at 33~GHz.

We can also use Eq. 13 by \citet{2012murphy1} to estimate the SFR from mid-infrared luminosity; instead of 24~$\mu$m, we will use the 22~$\mu$m value from the Wide-field Infrared Survey Explorer (WISE). However, it is worth noting that the AGN also contributes to the mid-infrared luminosity, and when polar dust is present, it can produce a mid-infrared excess, mimicking observational signatures of star formation \citep{2022jarvela1}. Keeping that in mind, interestingly, estimating the SFR from the 22$\mu$m luminosity gives exactly the same value of $\sim60~M_\odot$~yr$^{-1}$, but in this case, due to the resolution of WISE, for the whole galaxy. This is not an unreasonable value, especially for NLS1s, which are known to exhibit enhanced star formation \citep{2010sani1}. Alternative ways to assess the strength of star formation are the infrared colours and the $q22$ parameter. A very red W3-W4 colour (12$\mu$m$-$22$\mu$m), exceeding 2.5, indicates a strong star formation contribution to the infrared emission \citep{2015caccianiga1}, since the emission originating from star formation increases faster relative to the emission of AGN origin. Another parameter they suggest is $q22$, defined as log($S_{22~\mu \mathrm{m}}$ / $S_{1.4\mathrm{~GHz}}$). This parameter is analogous to the classical far-infrared - radio correlation parameter, but uses the WISE 22~$\mu$m band instead of far-infrared measurements. According to them, $q22$ exceeding unity can be a sign of enhanced star formation. For J1228+5017, W3-W4 = 2.45, and $q22$ = 1.9, indicating possibly enhanced star formation in the host.

All these proxies use the 22~$\mu$m measurement, so the question of what fraction of this emission comes from the AGN versus star formation remains. As mentioned earlier, AGN polar dust can enhance mid-infrared emission, and especially so in sources with strong narrow-line region outflows \citep{2022lyu1}. Indeed, J1228+5017's [O~III]$\lambda\lambda$4959,5007 emission lines exhibit a strong, high-velocity outflow \citep[see Fig.~2 in][]{2025crepaldi1}. Two wing components are required to properly model the outflow, with the higher-velocity component showing a ($-$371 $\pm$ 35) km~s$^{-1}$ blueshift. The main mechanisms to drive these outflows are the accretion disk winds \citep{2000proga1} and (relativistic) jets \citep{2001tadhunter1, 2008nesvabda1, 2016berton2}. In the case of J1228+5017, no persistent jets seem to be present, so the NLR outflow is likely driven by radiation pressure induced by the high Eddington ratio. All things considered, the current data, unfortunately, do not allow us to determine the origin of the 22~$\mu$m emission. %In addition to driving NLR outflows, jets can heat up dust grains and cause a mid-infrared excess similar to that in sources where radiation pressure drives the outflowing polar dust component \citep{2024haidar1}. So, 

However, other properties of our JVLA data do not indicate a star formation origin for the high-frequency radio emission. As can be seen in the light curves in Fig.~\ref{fig:j1228lc} and the spectra in Fig.~\ref{fig:j1228spect}, the K and Ka band emission seems to be variable. As we do not expect to see variability from star formation at these timescales, it is safe to assume that at least the varying fraction of the flux density is of AGN origin. Thus, if star formation is not likely to dominate the highest observed frequencies, we need to turn our attention back to AGN-produced emission. In the absence of extended optically thin emission, detecting the optically thick jet base can yield a flat spectrum \citep{2019panessa1}. However, we would have expected to detect it also in the mas-resolution VLBA observations at 15~GHz. The 1$\sigma$ rms of those observations was 58~$\mu$Jy, and thus any possible jet base emission was probably under the detection threshold \citep{2024jarvela1}. 

Another component that should start to become detectable toward higher radio frequencies is the tentative radio corona \citep{2008laor1}. Its possible physical connection to the jet base is still unclear, but for clarity, we will discuss it as a separate component. The radio corona is thought to be magnetically heated and, at its smallest scales at hundreds of GHz, cospatial with the X-ray corona, with a size around some tens of $r_\mathrm{g}$ --- similarly to coronally active stars. These stars, as well as several non-jetted AGN, follow the G\"udel-Benz relation: $L_\mathrm{R}/L_\mathrm{X} \sim 10^{-5}$ \citep{guedel93}. We estimate the X-ray luminosity from the \textit{Swift} observations assuming 0.03~cps between 0.3-10~keV and a photon index of 2.5. For radio, we use the earlier calculated $L_\mathrm{33~GHz}$ to calculate $\nu L_\nu$. This yields $L_\mathrm{R}/L_\mathrm{X} = 4\times 10^{-5}$, being consistent with the G\"udel-Bentz relation. Assuming equipartition and a power-law electron energy distribution with $p$=2, we can estimate the size of the radio-emitting region using Eq. 22 by \citet{2008laor1}:
\begin{equation}
    R^\mathrm{RS}_\mathrm{pc} = 0.47 L_{30}^{0.4} L_{46}^{0.1}\nu_\mathrm{GHz}^{-1},
\end{equation}
where $R^\mathrm{RS}_\mathrm{pc}$ is the size of the radio sphere in pc, $L_{30}$ is $L_\nu$/10$^{30}$~erg~s$^{-1}$~Hz$^{-1}$, $L_{46}$ is $L_\mathrm{bol}/10^{46}$ and $\nu_\mathrm{GHz}$ is the observed frequency in GHz. To estimate $L_{30}$, we use $L_\mathrm{33~GHz}$, and for the bolometric luminosity, we use the value derived from the monochromatic 5100~Å luminosity by \citet{2025crepaldi1}. J1228+5017 does not appear to host a persistent jet, and based on the multiwavelength observations so far, the extreme radio variability does not seem to have an optical counterpart; therefore, we assume that the 5100~Å luminosity is not contaminated by this activity. With these values, we obtain $R^\mathrm{RS}_\mathrm{pc}$ = 0.006~pc or 6~mpc, which corresponds to $\sim$1.4 $\times$ 10$^4~r_\mathrm{g}$, or around 7 light-days. Taken at face value, this size is likely to be too large to be related to the X-ray or the radio corona. However, it is consistent with the size of the emitting region derived from the variability timescale, $r <$ 0.03 $\times~\delta$, assuming $\delta >$ 100, but it should be noted that these estimates are based on very different assumptions. Both assume equipartition, but flare timescales assume relativistically boosted emission, whereas the coronal emission is expected to be close to isotropic. This could be a coincidence, or it could point to the low- and flaring-state emission originating from the same region.

% here could also estimate the B from timescales (2008 Laor Eq. 31)

% using Gamma = 2.5 and cps = 0.03, FLUX: PIMMS predicts a flux ( 0.300- 10.000keV) of 8.691E-13 ergs/cm^2/s
%Unabsorbed FLUX: PIMMS predicts an unabsorbed flux ( 0.300- 10.000keV) of 9.335E-13 ergs/cm^2/s
%L_X = 1.5 * 10^44 erg/s
%\nu L_R = 5.64*10^(39) erg/s (using (105.7 $\pm$ 19.1)~$\mu$Jy)

\subsection{Lack of multiwavelength trends}

Independently of the origin of the radio emission, the jet (base) or the corona, in several cases, we would expect to see correlated multiwavelength variability induced by the couplings between the accretion disk, the X-ray corona, and the radio corona or the jet. In another similar source, SDSS J164100.10+345452.7, X-rays brightened after a 37~GHz flare, possibly suggesting a connection between the two bands \citep{2023romano1}. However, in the case of J1228+5017, based on the multiwavelength light curves in Fig.~\ref{fig:j1228lc}, no obvious correlated variability can be seen between the 37~GHz flares and the other studied bands. There are a few possible explanations for this.

First, the cadence of the \textit{Swift} observations might be too low to catch the trends seen in the 37~GHz light curve. Assuming that the flare is quasi-simultaneous in different bands, and given the extremely fast timescales in radio, the X-ray observations would likely need to be simultaneous with the radio to catch the X-ray flare. On the other hand, in the case of SDSS J164100.10+345452.7, the X-ray emission stayed elevated for 10-20 days after the radio flare \citep{2023romano1}. This might point to a different origin or possibly environment between the two sources. Another option is that the flares at shorter wavelengths are too short to catch with the long \textit{Swift} exposures, yielding averages with quite similar, low cps values. Considering the faintness of the source, higher-sensitivity observations would be required to achieve higher temporal resolution. Third, it is also possible that there is a delay between the short-wavelength and radio emission, and the high-activity period in the optical-to-X-ray bands did not occur during our observations. Lastly, the radio flares may not always, or in this source, have multiwavelength counterparts. In blazars, orphan flares are seen over the electromagnetic spectrum, including radio frequencies \citep{2015hughes1, 2022wang1}. However, it is hard to say whether the comparison with J1228+5017 makes sense when the source does not seem to host standard relativistic jets.

\subsection{Nature of the flares} \label{sec:natureofflares}

The derived timescales are upper limits, but the flares cannot be arbitrarily short and pulse-like either; otherwise, they would not result in detections at MRO, and these pulses would also be seen in the temporally resolved MRO data. Unfortunately, due to sensitivity limitations, the MRO observations cannot be split into shorter integrations to study possible flux density changes within one $\sim$1800~s integration.

The most extraordinary --- and constraining -- properties of these flares are their extremely short rise and decay timescales coupled with very high variability amplitude. Such timescales have not been seen in AGN before; excluding sources likely affected by interstellar scintillation, the blazar S5 0716+714 might come closest. Based on intraday variability, the $T_\mathrm{b, var}$ of S5 0716+714 has been estimated to possibly be as high as $\sim$10$^{17}$~K \citep[e.g.,][]{1996wagner1}, but the inferred rise and decay timescales are still of the order of days, and thus much longer than what is seen in J1228+5017 \citep{2006ostorero1, 2013rani1}. With this combination of properties, these extreme flares do not seem to have similar counterparts in the literature, and they might represent a new kind of AGN variability.

The possibility that these flares could be caused by interstellar scintillation was rejected by \citet{2024jarvela1}, but is worth a brief discussion here. The Galactic coordinates of J1228+5017 are $l$ = 132.0, $b$ = +66.4, so the column density of free electrons can be expected to be very low \citep{2019hancock1}. The observing frequency of 37~GHz is also well in the weak-scattering regime, which usually produces amplitude variability of only 1--2\%. Moreover, to undergo scattering, the angular diameter of the source has to be approximately the same size or smaller than the first Fresnel zone of the scattering screen \citep{2008lazio1}. At 37~GHz, this is $\sim$10~$\mu$as; based on VLBA observations of these sources, no mas-scale components were detected.

\subsection{Origin of the flares}

With the currently available data, we cannot conclusively determine the physical origin of the radio flares. Revisiting the three hypotheses presented by \citet{2024jarvela1}, based on the timescales derived here, we can safely rule out the jet/jet base interaction with a BLR cloud/star, which is predicted to produce flares lasting 1-2 days. Next, we will discuss one of these hypotheses --- magnetic reconnection --- as a possible physical origin of these flares.

One viable explanation is magnetic eruptions near the black hole. A near-Eddington accretion flow carries magnetic loops of alternating polarity onto the horizon, where each loop is annihilated by the next one of opposite sign \citep{parfrey2015, chashkina2021}.
With $r_g = GM/c^{2} \approx 1.3\times10^{12}~\mathrm{cm}$ and $\dot M \approx L_\mathrm{Edd}/0.1c^{2} \approx 10^{25}~\mathrm{g}~\mathrm{s}^{-1}$, the horizon mass density $\rho \approx \dot M/4\pi r_g^{2}c \approx 2\times10^{-11}~\mathrm{g}~\mathrm{cm}^{-3}$ lets a loop saturate at an equipartition magnetic field strength of $B_\mathrm{H} \approx (8\pi\rho c^{2})^{1/2} \approx 6\times10^{5}~\mathrm{G}$, so a loop of size $h$ stores
\begin{equation}
  E_\mathrm{p} \approx \frac{B_\mathrm{H}^{2}}{8\pi}~h^{3}
               \approx 10^{48}\left(\frac{h}{3~r_g}\right)^{3}~\mathrm{erg} . \label{eq:Ep-summary}
\end{equation}
A burst radiates an isotropic-equivalent $E_\mathrm{rad} \sim 10^{45}$--$10^{47}~\mathrm{erg}$ per hour ($\nu L_\nu \approx 4.5\times10^{43}~\mathrm{erg}~\mathrm{s}^{-1}$ at the $0.6~$Jy peak), less than what one loop stores even without beaming. The loops also arrive on the inflow time of the magnetically elevated inner disk, $t_\mathrm{in} \approx$ days \citep[e.g.,][]{begelman2007}, matching the days-long intervals between bursts.

%\citep{granot2011}.
Reconnection between the loops can eject the annihilated flux into the polar cavity. The cavity contains mostly diluted $e^\pm$ plasma created by a pair cascade in the corona \citep[e.g.,][]{nattila2024}, so the region has a high magnetization $\sigma = B^{2}/4\pi n_\pm m_e c^{2} \gg 1$ and accelerates impulsively to $\Gamma \gg 1$. An observer on the axis sees $\delta \approx 2\Gamma$; $\delta \approx 140$ brings the intrinsic brightness temperature of the $1.1~$h rise, $T_\mathrm{b,var}/\delta^{3} \approx 10^{12}~\mathrm{K}$ ($3\times10^{10}~\mathrm{K}$ for the $8~$h burst duration), to the inverse-Compton limit \citep{kellermann1969}. Induced Compton scattering absorbs radiation of this brightness temperature in any plasma it crosses: the corona and the disk wind are opaque to the radio emission, and the cavity pairs become transparent only far away, beyond $r \gtrsim 10^{3}~r_g \approx 10^{15}~\mathrm{cm}$, so the shell must emit the burst beyond this radius and it can be viewed along the axis through a channel free of broad-line clouds.
%\citep{lyubarsky2008b}:

The shell can dissipate in several ways. Internal reconnection and turbulence between its stripes of alternating polarity energise the pairs to $\gamma \sim \sigma$; as they cool, pairs with $\gamma \approx 750$ ($\gamma \approx 15$ for the $8~$h burst) would emit at the $37~$GHz band at $r \sim 4\times10^{7}~r_g$ ($1.6\times10^{4}~r_g$), where the comoving field is $B' \approx 2\times10^{-4}~\mathrm{G}$ ($0.5~\mathrm{G}$). This synchrotron emission is broadband and linearly polarised up to $\sim 70\%$. It could be accompanied by a simultaneous hard X-ray flare from the energised pairs \citep[e.g.,][]{nattila2021}.

Alternatively, the shell keeps accelerating as $\Gamma \propto r^{1/3}$ to $\Gamma \sim 10^{3}$ and drives a magnetised shock into the ion-loaded disk wind at $r \gtrsim 10^{4}~r_g$, where the reduced wind magnetisation $\sigma_\mathrm{w} \sim 0.2$ lets the synchrotron maser radiate the shock precursor at an efficiency of $10^{-3}$ \citep{sironi2021}. This emission is coherent, linearly polarised (in the X mode), and narrow-band ($\Delta\nu/\nu \approx 0.6$), which reproduces the $15~$GHz non-detection, and drifts downward in frequency as the shell decelerates in the wind. Simultaneous multi-frequency and polarimetric radio coverage of one burst, together with a hard X-ray ($>$50~keV) exposure, could distinguish the two channels. \\

Another observable expected to be associated with magnetic reconnection events in AGN coronae is the Neupert effect \citep[e.g.,][]{neupert68, guedel02, 2008laor1}, whereby the light curves in both hard X-rays ($L_{\rm HX}(t)$) and radio ($L_{\rm R}(t)$) are correlated with the time derivative of the light curve in soft X-rays ($L_{\rm SX}(t)$),
\begin{equation}
  L_\mathrm{HX}(t) \propto L_\mathrm{R}(t) \propto \frac{d}{dt}L_\mathrm{SX}(t) , \label{eq:neupert}
\end{equation}
as observed from the solar corona during some solar flares. In the solar scenario, a stream of charged particles is accelerated along a closed magnetic field loop and rejoins the chromosphere at the loop footprint \citep[e.g.,][]{holman16}. Some of these particles may escape the flow as a coronal mass ejection (CME). The hard X-ray and radio flares are attributed to the instantaneous nonthermal bremsstrahlung and synchrotron emission of the downward-streaming particles. The more slowly variable soft X-rays are attributed primarily to thermal bremsstrahlung as the particle stream deposits the bulk of its energy into the foot of the loop, heating the chromospheric plasma and evaporating it into the corona. \citet{vardi26} report the observation of a solar-like AGN flare event in NGC 3783, wherein the hard and soft X-ray variability appears consistent with the Neupert effect from a magnetic reconnection event, and they suggest that an ultra-fast outflow observed by \citet{gu25} during the same flare event is a CME-like ejection from the magnetic loop stream. They also find a trailing rise in the UV, which they attribute to hard X-ray photons from the flare illuminating the accretion disk and enhancing its UV emission. 

With the present data, we cannot confirm or reject such an event for J1228+5017; while we do not see corresponding X-ray or UV enhancement in the light curves (Fig.~\ref{fig:j1228lc}), it may simply be that we do not have the necessary observing cadence. On the other hand, the G\"udel-Benz relation should also hold during CMEs, suggesting that a 10$^4$~Jy X-ray flare should accompany a 1~Jy radio flare; this kind of X-ray flare would surely have been picked up by X-ray monitoring instruments, such as the \textit{Swift}/Burst Alert Telescope \citep[BAT,][]{Barthelmy2005:BAT} or the Monitor of All-sky X-ray Image \citep[MAXI,][]{MAXI:Matsuoka2009PASJ}. Thus, it does not seem likely that these flares could be attributed to CME-like phenomena.

%At a minimum, testing for the Neupert effect in J1228+5017 will require dedicated, simultaneous radio and X-ray observations to capture both rapid radio and hard X-ray flares as well as a gradual soft X-ray flare.

\section{Conclusions} \label{sec:summary}

In this paper, we investigated the multiwavelength behaviour of the NLS1 galaxy J1228+5017 during a period of high activity at 37~GHz. We combined data from MRO, OVRO, JVLA, and \textit{Swift}/XRT and UVOT to further constrain the observational properties and the physical origin of these extreme high-frequency radio flares. Our conclusions are the following:

i) The extremely short inferred timescale ($e$-folding time of 1.1~h) coupled with the high, Jy-level amplitude variability and the absence of persistent relativistic jets likely rules out known AGN radio variability mechanisms, strongly suggesting that alternative explanations are required.

ii) The lack of multiwavelength trends might indicate that these flares are exclusively detectable in the radio regime. Alternatively, their higher-energy counterparts might occur at bands not probed here, such as the hard X-ray band, as suggested by the magnetic reconnection scenario.

iii) The emitting region size, a few mpc, seems to be inconsistent with the expected size of the radio corona at GHz frequencies. Interestingly, the sizes derived from the variability timescale and the low-state flux density are similar, possibly indicating that low- and high-state emission originates from the same region.

iv) Radio emission powered by magnetic eruptions stays a viable hypothesis for the physical origin of the flares. This would indicate high magnetic activity in these sources, as predicted by simulations of near-Eddington accreting supermassive black holes.

While these new data further constrain the properties of the extreme radio flares, their physical origin remains unclear. Targeted observations will be needed to test, for example, the predictions of the magnetic reconnection hypothesis. Future observations should focus on measuring the polarisation of the radio emission as well as obtaining high-time-resolution broadband dynamic radio spectra. It could provide a direct test between broadband synchrotron variability and the narrow-band, frequency-drifting emission expected in the synchrotron-maser scenario. These kinds of observations would tremendously help us narrow down the options. Furthermore, simultaneous radio and hard X-ray observations would directly test the magnetic eruption scenario. Due to the sporadic nature of these flares and their short timescales, these observations will not be easy, but considering that we might be witnessing a totally new type of AGN variability, they will be worth it.

\section*{Acknowledgements}

We want to thank an MRO observer, I. Syrjärinne, for obtaining some of the MRO data presented here.

JN acknowledges funding by the Research Council of Finland grant No. 372879 (OVERFLOW).

This publication makes use of data obtained at Metsähovi Radio Observatory, operated by Aalto University in Finland.

The National Radio Astronomy Observatory is a facility of the National Science Foundation operated under cooperative agreement by Associated Universities, Inc. CIRADA is funded by a grant from the Canada Foundation for Innovation 2017 Innovation Fund (Project 35999), as well as by the Provinces of Ontario, British Columbia, Alberta, Manitoba and Quebec.

This research has made use of data from the OVRO 40-m monitoring program \citep{2011richards1}, supported by private funding from the California Institute of Technology and the Max Planck Institute for Radio Astronomy, and by NASA grants NNX08AW31G, NNX11A043G, and NNX14AQ89G and NSF grants AST-0808050 and AST- 1109911.

This research has made use of the NASA/IPAC Extragalactic Database (NED), which is operated by the Jet Propulsion Laboratory, California Institute of Technology, under contract with the National Aeronautics and Space Administration. This research has made use of the SIMBAD database, operated at CDS, Strasbourg, France.

No AI tools were used in the creation of this manuscript.

%%%%%%%%%%%%%%%%%%%%%%%%%%%%%%%%%%%%%%%%%%%%%%%%%%
\section*{Data Availability}

The JVLA (Legacy ID: AJ442) data are publicly available in the NRAO Data Archive: https://data.nrao.edu. The MRO data will be made available via CDS. The OVRO 40m light curve underlying this article will be shared on reasonable request to the OVRO 40m Monitoring collaboration. The data underlying this article are publicly available from the \textit{Swift} archive.  



\bibliographystyle{aa}
\bibliography{artikkeli} % if your bibtex file is called example.bib

\begin{thebibliography}{89}
\expandafter\ifx\csname natexlab\endcsname\relax\def\natexlab#1{#1}\fi

\bibitem[{{Abdo} {et~al.}(2009){Abdo}, {Ackermann}, {Ajello}, {Axelsson}, {Baldini}, {Ballet}, {Barbiellini}, {Bastieri}, {Battelino}, {Baughman}, {Bechtol}, {Bellazzini}, {Bloom}, {Bonamente}, {Borgland}, {Bregeon}, {Brez}, {Brigida}, {Bruel}, {Caliandro}, {Cameron}, {Caraveo}, {Casandjian}, {Cavazzuti}, {Cecchi}, {Chekhtman}, {Cheung}, {Chiang}, {Ciprini}, {Claus}, {Cohen-Tanugi}, {Collmar}, {Conrad}, {Costamante}, {Dermer}, {de Angelis}, {de Palma}, {Digel}, {Silva}, {Drell}, {Dubois}, {Dumora}, {Farnier}, {Favuzzi}, {Focke}, {Foschini}, {Frailis}, {Fuhrmann}, {Fukazawa}, {Funk}, {Fusco}, {Gargano}, {Gehrels}, {Germani}, {Giebels}, {Giglietto}, {Giordano}, {Giroletti}, {Glanzman}, {Grenier}, {Grondin}, {Grove}, {Guillemot}, {Guiriec}, {Hanabata}, {Harding}, {Hartman}, {Hayashida}, {Hays}, {Hughes}, {J{\'o}hannesson}, {Johnson}, {Johnson}, {Johnson}, {Kamae}, {Katagiri}, {Kataoka}, {Kerr}, {Kn{\"o}dlseder}, {Kuehn}, {Kuss}, {Lande}, {Latronico}, {Lemoine-Goumard}, {Longo}, {Loparco}, {Lott}, {Lovellette},
  {Lubrano}, {Madejski}, {Makeev}, {Max-Moerbeck}, {Mazziotta}, {McConville}, {McEnery}, {Meurer}, {Michelson}, {Mitthumsiri}, {Mizuno}, {Monte}, {Monzani}, {Morselli}, {Moskalenko}, {Murgia}, {Nolan}, {Norris}, {Nuss}, {Ohsugi}, {Omodei}, {Orlando}, {Ormes}, {Paneque}, {Panetta}, {Parent}, {Pavlidou}, {Pearson}, {Pepe}, {Pesce-Rollins}, {Piron}, {Porter}, {Rain{\`o}}, {Rando}, {Razzano}, {Readhead}, {Reimer}, {Reimer}, {Reposeur}, {Richards}, {Ritz}, {Rodriguez}, {Romani}, {Ryde}, {Sadrozinski}, {Sambruna}, {Sanchez}, {Sander}, {Parkinson}, {Scargle}, {Schalk}, {Sgr{\`o}}, {Smith}, {Spandre}, {Spinelli}, {Starck}, {Stevenson}, {Strickman}, {Suson}, {Tagliaferri}, {Takahashi}, {Tanaka}, {Thayer}, {Thompson}, {Tibaldo}, {Tibolla}, {Torres}, {Tosti}, {Tramacere}, {Uchiyama}, {Usher}, {Vilchez}, {Vitale}, {Waite}, {Winer}, {Wood}, {Ylinen}, {Zensus}, {Ziegler}, {Fermi/LAT Collaboration}, {Ghisellini}, {Maraschi}, {Tavecchio}, \& {Angelakis}}]{2009abdo2}
{Abdo}, A.~A., {Ackermann}, M., {Ajello}, M., {et~al.} 2009, \apj, 699, 976

\bibitem[{{Arnaud}(1996)}]{Arnaud1996:xspec}
{Arnaud}, K.~A. 1996, in Astronomical Society of the Pacific Conference Series, Vol. 101, Astronomical Data Analysis Software and Systems V, ed. G.~H. {Jacoby} \& J.~{Barnes}, 17

\bibitem[{{Baars} {et~al.}(1977){Baars}, {Genzel}, {Pauliny-Toth}, \& {Witzel}}]{1977baars1}
{Baars}, J.~W.~M., {Genzel}, R., {Pauliny-Toth}, I.~I.~K., \& {Witzel}, A. 1977, \aap, 61, 99

\bibitem[{{Barthelmy} {et~al.}(2005){Barthelmy}, {Barbier}, {Cummings}, {Fenimore}, {Gehrels}, {Hullinger}, {Krimm}, {Markwardt}, {Palmer}, {Parsons}, {Sato}, {Suzuki}, {Takahashi}, {Tashiro}, \& {Tueller}}]{Barthelmy2005:BAT}
{Barthelmy}, S.~D., {Barbier}, L.~M., {Cummings}, J.~R., {et~al.} 2005, \ssr, 120, 143

\bibitem[{{Begelman} \& {Pringle}(2007)}]{begelman2007}
{Begelman}, M.~C. \& {Pringle}, J.~E. 2007, \mnras, 375, 1070

\bibitem[{{Berton} {et~al.}(2020{\natexlab{a}}){Berton}, {Bj{\"o}rklund}, {L{\"a}hteenm{\"a}ki}, {Congiu}, {J{\"a}rvel{\"a}}, {Terreran}, \& {La Mura}}]{2020berton1}
{Berton}, M., {Bj{\"o}rklund}, I., {L{\"a}hteenm{\"a}ki}, A., {et~al.} 2020{\natexlab{a}}, Contributions of the Astronomical Observatory Skalnate Pleso, 50, 270

\bibitem[{{Berton} {et~al.}(2017){Berton}, {Foschini}, {Caccianiga}, {Ciroi}, {Congiu}, {Cracco}, {Frezzato}, {La Mura}, \& {Rafanelli}}]{2017berton1}
{Berton}, M., {Foschini}, L., {Caccianiga}, A., {et~al.} 2017, Frontiers in Astronomy and Space Sciences, 4, 8

\bibitem[{{Berton} {et~al.}(2016){Berton}, {Foschini}, {Ciroi}, {Cracco}, {La Mura}, {Di Mille}, \& {Rafanelli}}]{2016berton2}
{Berton}, M., {Foschini}, L., {Ciroi}, S., {et~al.} 2016, \aap, 591, A88

\bibitem[{{Berton} {et~al.}(2020{\natexlab{b}}){Berton}, {J{\"a}rvel{\"a}}, {Crepaldi}, {L{\"a}hteenm{\"a}ki}, {Tornikoski}, {Congiu}, {Kharb}, {Terreran}, \& {Vietri}}]{2020berton2}
{Berton}, M., {J{\"a}rvel{\"a}}, E., {Crepaldi}, L., {et~al.} 2020{\natexlab{b}}, \aap, 636, A64

\bibitem[{{Blanchard} {et~al.}(2017){Blanchard}, {Nicholl}, {Berger}, {Guillochon}, {Margutti}, {Chornock}, {Alexander}, {Leja}, \& {Drout}}]{2017blanchard1}
{Blanchard}, P.~K., {Nicholl}, M., {Berger}, E., {et~al.} 2017, \apj, 843, 106

\bibitem[{{Boroson} \& {Green}(1992)}]{1992boroson1}
{Boroson}, T.~A. \& {Green}, R.~F. 1992, \apjs, 80, 109

\bibitem[{{Burrows} {et~al.}(2005){Burrows}, {Hill}, {Nousek}, {Kennea}, {Wells}, {Osborne}, {Abbey}, {Beardmore}, {Mukerjee}, {Short}, {Chincarini}, {Campana}, {Citterio}, {Moretti}, {Pagani}, {Tagliaferri}, {Giommi}, {Capalbi}, {Tamburelli}, {Angelini}, {Cusumano}, {Br{\"a}uninger}, {Burkert}, \& {Hartner}}]{Burrows2005:XRT}
{Burrows}, D.~N., {Hill}, J.~E., {Nousek}, J.~A., {et~al.} 2005, \ssr, 120, 165

\bibitem[{{Caccianiga} {et~al.}(2015){Caccianiga}, {Ant{\'o}n}, {Ballo}, {Foschini}, {Maccacaro}, {Della Ceca}, {Severgnini}, {March{\~a}}, {Mateos}, \& {Sani}}]{2015caccianiga1}
{Caccianiga}, A., {Ant{\'o}n}, S., {Ballo}, L., {et~al.} 2015, \mnras, 451, 1795

\bibitem[{{Chashkina} {et~al.}(2021){Chashkina}, {Bromberg}, \& {Levinson}}]{chashkina2021}
{Chashkina}, A., {Bromberg}, O., \& {Levinson}, A. 2021, \mnras, 508, 1241

\bibitem[{{Cracco} {et~al.}(2016){Cracco}, {Ciroi}, {Berton}, {Di Mille}, {Foschini}, {La Mura}, \& {Rafanelli}}]{2016cracco1}
{Cracco}, V., {Ciroi}, S., {Berton}, M., {et~al.} 2016, \mnras, 462, 1256

\bibitem[{{Crepaldi} {et~al.}(2025){Crepaldi}, {Berton}, {Dalla Barba}, {La Mura}, {J{\"a}rvel{\"a}}, {Vietri}, \& {Ciroi}}]{2025crepaldi1}
{Crepaldi}, L., {Berton}, M., {Dalla Barba}, B., {et~al.} 2025, \aap, 696, A74

\bibitem[{{Dignan} {et~al.}(2025){Dignan}, {Murphy}, {Mason}, {Eibensteiner}, {Hensley}, {Jim{\'e}nez-Andrade}, {Linden}, {Dicker}, {Dong}, {Momjian}, {Romero}, {Schinnerer}, \& {Turner}}]{2025dignan1}
{Dignan}, A., {Murphy}, E.~J., {Mason}, B., {et~al.} 2025, \apj, 988, 216

\bibitem[{{Du} {et~al.}(2018){Du}, {Zhang}, {Wang}, {Huang}, {Zhang}, {Lu}, {Hu}, {Li}, {Bai}, {Bian}, {Yuan}, {Ho}, {Wang}, \& {SEAMBH Collaboration}}]{2018du1}
{Du}, P., {Zhang}, Z.-X., {Wang}, K., {et~al.} 2018, \apj, 856, 6

\bibitem[{{Frederick} {et~al.}(2021){Frederick}, {Gezari}, {Graham}, {Sollerman}, {van Velzen}, {Perley}, {Stern}, {Ward}, {Hammerstein}, {Hung}, {Yan}, {Andreoni}, {Bellm}, {Duev}, {Kowalski}, {Mahabal}, {Masci}, {Medford}, {Rusholme}, {Smith}, \& {Walters}}]{2021frederick1}
{Frederick}, S., {Gezari}, S., {Graham}, M.~J., {et~al.} 2021, \apj, 920, 56

\bibitem[{{Gehrels} {et~al.}(2004){Gehrels}, {Chincarini}, {Giommi}, {Mason}, {Nousek}, {Wells}, {White}, {Barthelmy}, {Burrows}, {Cominsky}, {Hurley}, {Marshall}, {M{\'e}sz{\'a}ros}, {Roming}, {Angelini}, {Barbier}, {Belloni}, {Campana}, {Caraveo}, {Chester}, {Citterio}, {Cline}, {Cropper}, {Cummings}, {Dean}, {Feigelson}, {Fenimore}, {Frail}, {Fruchter}, {Garmire}, {Gendreau}, {Ghisellini}, {Greiner}, {Hill}, {Hunsberger}, {Krimm}, {Kulkarni}, {Kumar}, {Lebrun}, {Lloyd-Ronning}, {Markwardt}, {Mattson}, {Mushotzky}, {Norris}, {Osborne}, {Paczynski}, {Palmer}, {Park}, {Parsons}, {Paul}, {Rees}, {Reynolds}, {Rhoads}, {Sasseen}, {Schaefer}, {Short}, {Smale}, {Smith}, {Stella}, {Tagliaferri}, {Takahashi}, {Tashiro}, {Townsley}, {Tueller}, {Turner}, {Vietri}, {Voges}, {Ward}, {Willingale}, {Zerbi}, \& {Zhang}}]{Gehrels2004}
{Gehrels}, N., {Chincarini}, G., {Giommi}, P., {et~al.} 2004, \apj, 611, 1005

\bibitem[{{Goad} {et~al.}(2012){Goad}, {Korista}, \& {Ruff}}]{2012goad1}
{Goad}, M.~R., {Korista}, K.~T., \& {Ruff}, A.~J. 2012, \mnras, 426, 3086

\bibitem[{{Goodrich}(1989)}]{1989goodrich1}
{Goodrich}, R.~W. 1989, \apj, 342, 224

\bibitem[{{Gu} {et~al.}(2025){Gu}, {Fukumura}, {Kaastra}, {Eckart}, {Ballhausen}, {Behar}, {Diez}, {Guainazzi}, {Kallman}, {Kara}, {Li}, {Mehdipour}, {Mizumoto}, {Ogawa}, {Panagiotou}, {Signorini}, {Tanimoto}, {Zhao}, {Noda}, {Miller}, \& {Yamada}}]{gu25}
{Gu}, L., {Fukumura}, K., {Kaastra}, J., {et~al.} 2025, \aap, 704, A146

\bibitem[{{G{\"u}del} {et~al.}(2002){G{\"u}del}, {Audard}, {Smith}, {Behar}, {Beasley}, \& {Mewe}}]{guedel02}
{G{\"u}del}, M., {Audard}, M., {Smith}, K.~W., {et~al.} 2002, \apj, 577, 371

\bibitem[{{Guedel} \& {Benz}(1993)}]{guedel93}
{Guedel}, M. \& {Benz}, A.~O. 1993, \apjl, 405, L63

\bibitem[{{Hancock} {et~al.}(2019){Hancock}, {Charlton}, {Macquart}, \& {Hurley-Walker}}]{2019hancock1}
{Hancock}, P.~J., {Charlton}, E.~G., {Macquart}, J.-P., \& {Hurley-Walker}, N. 2019, arXiv e-prints, arXiv:1907.08395

\bibitem[{{HI4PI Collaboration} {et~al.}(2016){HI4PI Collaboration}, {Ben Bekhti}, {Fl{\"o}er}, {Keller}, {Kerp}, {Lenz}, {Winkel}, {Bailin}, {Calabretta}, {Dedes}, {Ford}, {Gibson}, {Haud}, {Janowiecki}, {Kalberla}, {Lockman}, {McClure-Griffiths}, {Murphy}, {Nakanishi}, {Pisano}, \& {Staveley-Smith}}]{NH2016:HI4PI}
{HI4PI Collaboration}, {Ben Bekhti}, N., {Fl{\"o}er}, L., {et~al.} 2016, \aap, 594, A116

\bibitem[{{Holman}(2016)}]{holman16}
{Holman}, G.~D. 2016, Journal of Geophysical Research (Space Physics), 121, 11,667

\bibitem[{{Homan} {et~al.}(2021){Homan}, {Cohen}, {Hovatta}, {Kellermann}, {Kovalev}, {Lister}, {Popkov}, {Pushkarev}, {Ros}, \& {Savolainen}}]{2021homan1}
{Homan}, D.~C., {Cohen}, M.~H., {Hovatta}, T., {et~al.} 2021, \apj, 923, 67

\bibitem[{{Hovatta} {et~al.}(2009){Hovatta}, {Valtaoja}, {Tornikoski}, \& {L{\"a}hteenm{\"a}ki}}]{2009hovatta1}
{Hovatta}, T., {Valtaoja}, E., {Tornikoski}, M., \& {L{\"a}hteenm{\"a}ki}, A. 2009, \aap, 494, 527

\bibitem[{{Hughes} {et~al.}(2015){Hughes}, {Aller}, \& {Aller}}]{2015hughes1}
{Hughes}, P.~A., {Aller}, M.~F., \& {Aller}, H.~D. 2015, \apj, 799, 207

\bibitem[{{J{\"a}rvel{\"a}} {et~al.}(2021){J{\"a}rvel{\"a}}, {Berton}, \& {Crepaldi}}]{2021jarvela1}
{J{\"a}rvel{\"a}}, E., {Berton}, M., \& {Crepaldi}, L. 2021, Frontiers in Astronomy and Space Sciences, 8, 147

\bibitem[{{J{\"a}rvel{\"a}} {et~al.}(2022){J{\"a}rvel{\"a}}, {Dahale}, {Crepaldi}, {Berton}, {Congiu}, \& {Antonucci}}]{2022jarvela1}
{J{\"a}rvel{\"a}}, E., {Dahale}, R., {Crepaldi}, L., {et~al.} 2022, \aap, 658, A12

\bibitem[{{J{\"a}rvel{\"a}} {et~al.}(2017){J{\"a}rvel{\"a}}, {L{\"a}hteenm{\"a}ki}, {Lietzen}, {Poudel}, {Hein{\"a}m{\"a}ki}, \& {Einasto}}]{2017jarvela1}
{J{\"a}rvel{\"a}}, E., {L{\"a}hteenm{\"a}ki}, A., {Lietzen}, H., {et~al.} 2017, \aap, 606, A9

\bibitem[{{J{\"a}rvel{\"a}} {et~al.}(2024){J{\"a}rvel{\"a}}, {Savolainen}, {Berton}, {L{\"a}hteenm{\"a}ki}, {Kiehlmann}, {Hovatta}, {Varglund}, {Readhead}, {Tornikoski}, {Max-Moerbeck}, \& et~al.}]{2024jarvela1}
{J{\"a}rvel{\"a}}, E., {Savolainen}, T., {Berton}, M., {et~al.} 2024, \mnras, 532, 3069

\bibitem[{{Katebi}(2020)}]{2020katebi1}
{Katebi}, R. 2020, in American Astronomical Society Meeting Abstracts, Vol. 235, American Astronomical Society Meeting Abstracts \#235, 346.07

\bibitem[{{Kellermann} \& {Pauliny-Toth}(1969)}]{kellermann1969}
{Kellermann}, K.~I. \& {Pauliny-Toth}, I.~I.~K. 1969, \apjl, 155, L71

\bibitem[{{Kellermann} {et~al.}(1989){Kellermann}, {Sramek}, {Schmidt}, {Shaffer}, \& {Green}}]{1989kellermann1}
{Kellermann}, K.~I., {Sramek}, R., {Schmidt}, M., {Shaffer}, D.~B., \& {Green}, R. 1989, \aj, 98, 1195

\bibitem[{{Klein} {et~al.}(2018){Klein}, {Lisenfeld}, \& {Verley}}]{2018klein1}
{Klein}, U., {Lisenfeld}, U., \& {Verley}, S. 2018, \aap, 611, A55

\bibitem[{{Kollatschny} \& {Zetzl}(2011)}]{2011kollatschny1}
{Kollatschny}, W. \& {Zetzl}, M. 2011, \nat, 470, 366

\bibitem[{{Komossa} {et~al.}(2006){Komossa}, {Voges}, {Xu}, {Mathur}, {Adorf}, {Lemson}, {Duschl}, \& {Grupe}}]{2006komossa1}
{Komossa}, S., {Voges}, W., {Xu}, D., {et~al.} 2006, \aj, 132, 531

\bibitem[{{Komossa} {et~al.}(2018){Komossa}, {Xu}, \& {Wagner}}]{2018komossa1}
{Komossa}, S., {Xu}, D.~W., \& {Wagner}, A.~Y. 2018, \mnras, 477, 5115

\bibitem[{{L{\"a}hteenm{\"a}ki} {et~al.}(2017){L{\"a}hteenm{\"a}ki}, {J{\"a}rvel{\"a}}, {Hovatta}, {Tornikoski}, {Harrison}, {L{\'o}pez-Caniego}, {Max-Moerbeck}, {Mingaliev}, {Pearson}, {Ramakrishnan}, {Readhead}, {Reeves}, {Richards}, {Sotnikova}, \& {Tammi}}]{2017lahteenmaki1}
{L{\"a}hteenm{\"a}ki}, A., {J{\"a}rvel{\"a}}, E., {Hovatta}, T., {et~al.} 2017, \aap, 603, A100

\bibitem[{{L{\"a}hteenm{\"a}ki} {et~al.}(2018){L{\"a}hteenm{\"a}ki}, {J{\"a}rvel{\"a}}, {Ramakrishnan}, {Tornikoski}, {Tammi}, {Vera}, \& {Chamani}}]{2018lahteenmaki1}
{L{\"a}hteenm{\"a}ki}, A., {J{\"a}rvel{\"a}}, E., {Ramakrishnan}, V., {et~al.} 2018, \aap, 614, L1

\bibitem[{{L{\"a}hteenm{\"a}ki} {et~al.}(1999){L{\"a}hteenm{\"a}ki}, {Valtaoja}, \& {Wiik}}]{1999lahteenmaki1}
{L{\"a}hteenm{\"a}ki}, A., {Valtaoja}, E., \& {Wiik}, K. 1999, \apj, 511, 112

\bibitem[{{Laor} \& {Behar}(2008)}]{2008laor1}
{Laor}, A. \& {Behar}, E. 2008, \mnras, 390, 847

\bibitem[{{Lazio} {et~al.}(2008){Lazio}, {Ojha}, {Fey}, {Kedziora-Chudczer}, {Cordes}, {Jauncey}, \& {Lovell}}]{2008lazio1}
{Lazio}, T. J.~W., {Ojha}, R., {Fey}, A.~L., {et~al.} 2008, \apj, 672, 115

\bibitem[{{Lyu} \& {Rieke}(2022)}]{2022lyu1}
{Lyu}, J. \& {Rieke}, G.~H. 2022, \apjl, 940, L31

\bibitem[{{Marziani} {et~al.}(2025){Marziani}, {Garnica Luna}, {Floris}, {del Olmo}, {Deconto-Machado}, {Buendia-Rios}, {Negrete}, \& {Dultzin}}]{2025marziani1}
{Marziani}, P., {Garnica Luna}, K., {Floris}, A., {et~al.} 2025, Universe, 11, 69

\bibitem[{{Marziani} {et~al.}(2003){Marziani}, {Zamanov}, {Sulentic}, \& {Calvani}}]{2003marziani1}
{Marziani}, P., {Zamanov}, R.~K., {Sulentic}, J.~W., \& {Calvani}, M. 2003, \mnras, 345, 1133

\bibitem[{{Mathur}(2000)}]{2000mathur1}
{Mathur}, S. 2000, \mnras, 314, L17

\bibitem[{{Matsuoka} {et~al.}(2009){Matsuoka}, {Kawasaki}, {Ueno}, {Tomida}, {Kohama}, {Suzuki}, {Adachi}, {Ishikawa}, {Mihara}, {Sugizaki}, {Isobe}, {Nakagawa}, {Tsunemi}, {Miyata}, {Kawai}, {Kataoka}, {Morii}, {Yoshida}, {Negoro}, {Nakajima}, {Ueda}, {Chujo}, {Yamaoka}, {Yamazaki}, {Nakahira}, {You}, {Ishiwata}, {Miyoshi}, {Eguchi}, {Hiroi}, {Katayama}, \& {Ebisawa}}]{MAXI:Matsuoka2009PASJ}
{Matsuoka}, M., {Kawasaki}, K., {Ueno}, S., {et~al.} 2009, \pasj, 61, 999

\bibitem[{{Murphy} {et~al.}(2012){Murphy}, {Bremseth}, {Mason}, {Condon}, {Schinnerer}, {Aniano}, {Armus}, {Helou}, {Turner}, \& {Jarrett}}]{2012murphy1}
{Murphy}, E.~J., {Bremseth}, J., {Mason}, B.~S., {et~al.} 2012, \apj, 761, 97

\bibitem[{{N{\"a}ttil{\"a}}(2024)}]{nattila2024}
{N{\"a}ttil{\"a}}, J. 2024, Nature Communications, 15, 7026

\bibitem[{{Nesvadba} {et~al.}(2008){Nesvadba}, {Lehnert}, {De Breuck}, {Gilbert}, \& {van Breugel}}]{2008nesvabda1}
{Nesvadba}, N.~P.~H., {Lehnert}, M.~D., {De Breuck}, C., {Gilbert}, A.~M., \& {van Breugel}, W. 2008, \aap, 491, 407

\bibitem[{{Neupert}(1968)}]{neupert68}
{Neupert}, W.~M. 1968, \apjl, 153, L59

\bibitem[{{Nättilä} \& {Beloborodov}(2021)}]{nattila2021}
{Nättilä}, J. \& {Beloborodov}, A.~M. 2021, \apj, 921, 87

\bibitem[{{Olgu{\'\i}n-Iglesias} {et~al.}(2020){Olgu{\'\i}n-Iglesias}, {Kotilainen}, \& {Chavushyan}}]{2020olguiniglesias1}
{Olgu{\'\i}n-Iglesias}, A., {Kotilainen}, J., \& {Chavushyan}, V. 2020, \mnras, 492, 1450

\bibitem[{{Orban de Xivry} {et~al.}(2011){Orban de Xivry}, {Davies}, {Schartmann}, {Komossa}, {Marconi}, {Hicks}, {Engel}, \& {Tacconi}}]{2011orbandexivry1}
{Orban de Xivry}, G., {Davies}, R., {Schartmann}, M., {et~al.} 2011, \mnras, 417, 2721

\bibitem[{{Osterbrock} \& {Pogge}(1985)}]{1985osterbrock1}
{Osterbrock}, D.~E. \& {Pogge}, R.~W. 1985, \apj, 297, 166

\bibitem[{{Ostorero} {et~al.}(2006){Ostorero}, {Wagner}, {Gracia}, {Ferrero}, {Krichbaum}, {Britzen}, {Witzel}, {Nilsson}, {Villata}, {Bach}, {Barnaby}, {Bernhart}, {Carini}, {Chen}, {Chen}, {Ciprini}, {Crapanzano}, {Doroshenko}, {Efimova}, {Emmanoulopoulos}, {Fuhrmann}, {Gabanyi}, {Giltinan}, {Hagen-Thorn}, {Hauser}, {Heidt}, {Hojaev}, {Hovatta}, {Hroch}, {Ibrahimov}, {Impellizzeri}, {Ivanidze}, {Kachel}, {Kraus}, {Kurtanidze}, {L{\"a}hteenm{\"a}ki}, {Lanteri}, {Larionov}, {Lin}, {Lindfors}, {Munz}, {Nikolashvili}, {Nucciarelli}, {O'Connor}, {Ohlert}, {Pasanen}, {Pullen}, {Raiteri}, {Rector}, {Robb}, {Sigua}, {Sillanp{\"a}{\"a}}, {Sixtova}, {Smith}, {Strub}, {Takahashi}, {Takalo}, {Tapken}, {Tartar}, {Tornikoski}, {Tosti}, {Tr{\"o}ller}, {Walters}, {Wilking}, {Wills}, {Agudo}, {Aller}, {Aller}, {Angelakis}, {Klare}, {K{\"o}rding}, {Strom}, {Ter{\"a}sranta}, {Ungerechts}, \& {Vila-Vilar{\'o}}}]{2006ostorero1}
{Ostorero}, L., {Wagner}, S.~J., {Gracia}, J., {et~al.} 2006, \aap, 451, 797

\bibitem[{{Padovani}(2017)}]{2017padovani1}
{Padovani}, P. 2017, Nature Astronomy, 1, 0194

\bibitem[{{Panessa} {et~al.}(2019){Panessa}, {Baldi}, {Laor}, {Padovani}, {Behar}, \& {McHardy}}]{2019panessa1}
{Panessa}, F., {Baldi}, R.~D., {Laor}, A., {et~al.} 2019, Nature Astronomy, 3, 387

\bibitem[{{Parfrey} {et~al.}(2015){Parfrey}, {Giannios}, \& {Beloborodov}}]{parfrey2015}
{Parfrey}, K., {Giannios}, D., \& {Beloborodov}, A.~M. 2015, \mnras, 446, L61

\bibitem[{{Peterson}(2011)}]{2011peterson1}
{Peterson}, B.~M. 2011, ArXiv e-prints [\eprint[arXiv]{1109.4181}]

\bibitem[{{Proga} {et~al.}(2000){Proga}, {Stone}, \& {Kallman}}]{2000proga1}
{Proga}, D., {Stone}, J.~M., \& {Kallman}, T.~R. 2000, \apj, 543, 686

\bibitem[{{Rani} {et~al.}(2013){Rani}, {Krichbaum}, {Fuhrmann}, {B{\"o}ttcher}, {Lott}, {Aller}, {Aller}, {Angelakis}, {Bach}, {Bastieri}, {Falcone}, {Fukazawa}, {Gabanyi}, {Gupta}, {Gurwell}, {Itoh}, {Kawabata}, {Krips}, {L{\"a}hteenm{\"a}ki}, {Liu}, {Marchili}, {Max-Moerbeck}, {Nestoras}, {Nieppola}, {Quintana-Lacaci}, {Readhead}, {Richards}, {Sasada}, {Sievers}, {Sokolovsky}, {Stroh}, {Tammi}, {Tornikoski}, {Uemura}, {Ungerechts}, {Urano}, \& {Zensus}}]{2013rani1}
{Rani}, B., {Krichbaum}, T.~P., {Fuhrmann}, L., {et~al.} 2013, \aap, 552, A11

\bibitem[{{Readhead}(1994)}]{1994readhead1}
{Readhead}, A. C.~S. 1994, \apj, 426, 51

\bibitem[{{Richards} {et~al.}(2011){Richards}, {Max-Moerbeck}, {Pavlidou}, {King}, {Pearson}, {Readhead}, {Reeves}, {Shepherd}, {Stevenson}, {Weintraub}, {Fuhrmann}, {Angelakis}, {Zensus}, {Healey}, {Romani}, {Shaw}, {Grainge}, {Birkinshaw}, {Lancaster}, {Worrall}, {Taylor}, {Cotter}, \& {Bustos}}]{2011richards1}
{Richards}, J.~L., {Max-Moerbeck}, W., {Pavlidou}, V., {et~al.} 2011, \apjs, 194, 29

\bibitem[{{Romano} {et~al.}(2023){Romano}, {L{\"a}hteenm{\"a}ki}, {Vercellone}, {Foschini}, {Berton}, {Raiteri}, {Braito}, {Ciroi}, {J{\"a}rvel{\"a}}, {Baitieri}, {Varglund}, {Tornikoski}, \& {Suutarinen}}]{2023romano1}
{Romano}, P., {L{\"a}hteenm{\"a}ki}, A., {Vercellone}, S., {et~al.} 2023, \aap, 673, A85

\bibitem[{{Roming} {et~al.}(2005){Roming}, {Kennedy}, {Mason}, {Nousek}, {Ahr}, {Bingham}, {Broos}, {Carter}, {Hancock}, {Huckle}, {Hunsberger}, {Kawakami}, {Killough}, {Koch}, {McLelland}, {Smith}, {Smith}, {Soto}, {Boyd}, {Breeveld}, {Holland}, {Ivanushkina}, {Pryzby}, {Still}, \& {Stock}}]{Roming2005:UVOT}
{Roming}, P.~W.~A., {Kennedy}, T.~E., {Mason}, K.~O., {et~al.} 2005, \ssr, 120, 95

\bibitem[{{Sani} {et~al.}(2010){Sani}, {Lutz}, {Risaliti}, {Netzer}, {Gallo}, {Trakhtenbrot}, {Sturm}, \& {Boller}}]{2010sani1}
{Sani}, E., {Lutz}, D., {Risaliti}, G., {et~al.} 2010, \mnras, 403, 1246

\bibitem[{{Singh} \& {Meintjes}(2020)}]{2020singh1}
{Singh}, K.~K. \& {Meintjes}, P.~J. 2020, Astronomische Nachrichten, 341, 713

\bibitem[{{Sironi} {et~al.}(2021){Sironi}, {Plotnikov}, {N{\"a}ttil{\"a}}, \& {Beloborodov}}]{sironi2021}
{Sironi}, L., {Plotnikov}, I., {N{\"a}ttil{\"a}}, J., \& {Beloborodov}, A.~M. 2021, \prl, 127, 035101

\bibitem[{{Spergel} {et~al.}(2007){Spergel}, {Bean}, {Dor{\'e}}, {Nolta}, {Bennett}, {Dunkley}, {Hinshaw}, {Jarosik}, {Komatsu}, {Page}, {Peiris}, {Verde}, {Halpern}, {Hill}, {Kogut}, {Limon}, {Meyer}, {Odegard}, {Tucker}, {Weiland}, {Wollack}, \& {Wright}}]{2007spergel1}
{Spergel}, D.~N., {Bean}, R., {Dor{\'e}}, O., {et~al.} 2007, \apjs, 170, 377

\bibitem[{{Sulentic} {et~al.}(2000){Sulentic}, {Zwitter}, {Marziani}, \& {Dultzin-Hacyan}}]{2000sulentic1}
{Sulentic}, J.~W., {Zwitter}, T., {Marziani}, P., \& {Dultzin-Hacyan}, D. 2000, \apjl, 536, L5

\bibitem[{{Tadhunter} {et~al.}(2001){Tadhunter}, {Wills}, {Morganti}, {Oosterloo}, \& {Dickson}}]{2001tadhunter1}
{Tadhunter}, C., {Wills}, K., {Morganti}, R., {Oosterloo}, T., \& {Dickson}, R. 2001, \mnras, 327, 227

\bibitem[{{Teraesranta} {et~al.}(1998){Teraesranta}, {Tornikoski}, {Mujunen}, {Karlamaa}, {Valtonen}, {Henelius}, {Urpo}, {Lainela}, {Pursimo}, {Nilsson}, {Wiren}, {Laehteenmaeki}, {Korpi}, {Rekola}, {Heinaemaeki}, {Hanski}, {Nurmi}, {Kokkonen}, {Keinaenen}, {Joutsamo}, {Oksanen}, {Pietilae}, {Valtaoja}, {Valtonen}, \& {Koenoenen}}]{1998terasranta1}
{Teraesranta}, H., {Tornikoski}, M., {Mujunen}, A., {et~al.} 1998, \aaps, 132, 305

\bibitem[{{Valtaoja} {et~al.}(1999){Valtaoja}, {L{\"a}hteenm{\"a}ki}, {Ter{\"a}sranta}, \& {Lainela}}]{1999valtaoja1}
{Valtaoja}, E., {L{\"a}hteenm{\"a}ki}, A., {Ter{\"a}sranta}, H., \& {Lainela}, M. 1999, \apjs, 120, 95

\bibitem[{{Vardi} {et~al.}(2026){Vardi}, {Behar}, {Gu}, {Kaastra}, {Guainazzi}, {Mehdipour}, {Fukumura}, {Miller}, {Laor}, {Kara}, {Eckart}, {Mizumoto}, {Panagiotou}, {Li}, {Shoji}, {Signorini}, \& {Zhao}}]{vardi26}
{Vardi}, G., {Behar}, E., {Gu}, L., {et~al.} 2026, \apjl, 1007, L23

\bibitem[{{Varglund} {et~al.}(2023){Varglund}, {J{\"a}rvel{\"a}}, {Ciroi}, {Berton}, {Congiu}, {L{\"a}hteenm{\"a}ki}, \& {Di Mille}}]{2023varglund1}
{Varglund}, I., {J{\"a}rvel{\"a}}, E., {Ciroi}, S., {et~al.} 2023, \aap, 679, A32

\bibitem[{{Varglund} {et~al.}(2022){Varglund}, {J{\"a}rvel{\"a}}, {L{\"a}hteenm{\"a}ki}, {Berton}, {Ciroi}, \& {Congiu}}]{2022varglund1}
{Varglund}, I., {J{\"a}rvel{\"a}}, E., {L{\"a}hteenm{\"a}ki}, A., {et~al.} 2022, \aap, 668, A91

\bibitem[{{Vietri} {et~al.}(2022){Vietri}, {J{\"a}rvel{\"a}}, {Berton}, {Ciroi}, {Congiu}, {Chen}, \& {Di Mille}}]{2022vietri1}
{Vietri}, A., {J{\"a}rvel{\"a}}, E., {Berton}, M., {et~al.} 2022, \aap, 662, A20

\bibitem[{{Wagner} {et~al.}(1996){Wagner}, {Witzel}, {Heidt}, {Krichbaum}, {Qian}, {Quirrenbach}, {Wegner}, {Aller}, {Aller}, {Anton}, \& et~al.}]{1996wagner1}
{Wagner}, S.~J., {Witzel}, A., {Heidt}, J., {et~al.} 1996, \aj, 111, 2187

\bibitem[{{Wang} {et~al.}(2016){Wang}, {Du}, {Hu}, {Bai}, {Wang}, {Yi}, {Wang}, {Zhang}, {Xin}, {Lun}, {Chang}, \& {Fan}}]{2016wang1}
{Wang}, F., {Du}, P., {Hu}, C., {et~al.} 2016, \apj, 824, 149

\bibitem[{{Wang} {et~al.}(2022){Wang}, {Liu}, {Petropoulou}, {Oikonomou}, {Xue}, \& {Wang}}]{2022wang1}
{Wang}, Z.-R., {Liu}, R.-Y., {Petropoulou}, M., {et~al.} 2022, \prd, 105, 023005

\bibitem[{{Wilms} {et~al.}(2000){Wilms}, {Allen}, \& {McCray}}]{Wilms2000}
{Wilms}, J., {Allen}, A., \& {McCray}, R. 2000, \apj, 542, 914

\bibitem[{{Winkel} {et~al.}(2022){Winkel}, {Husemann}, {Davis}, {Smirnova-Pinchukova}, {Bennert}, {Combes}, {Gaspari}, {Jahnke}, {Neumann}, {O'Dea}, {P{\'e}rez-Torres}, {Singha}, {Tremblay}, \& {Rix}}]{2022winkel1}
{Winkel}, N., {Husemann}, B., {Davis}, T.~A., {et~al.} 2022, \aap, 663, A104

\bibitem[{{Yuan} {et~al.}(2008){Yuan}, {Zhou}, {Komossa}, {Dong}, {Wang}, {Lu}, \& {Bai}}]{2008yuan1}
{Yuan}, W., {Zhou}, H.~Y., {Komossa}, S., {et~al.} 2008, \apj, 685, 801

\end{thebibliography}

\begin{appendix}
\onecolumn

\section{Observation details}

The details of the \textit{Swift}/XRT observations are given in Table~\ref{tab:swift_xrt_log}

%\begin{table*} 	
 \begin{table}[ht!]
 \begin{center} 	
 \caption{{\it Swift}/XRT observation log: observing sequence, date (MJD of the middle of the observation), start and end times (UTC), and XRT exposure time.} 	
 \label{tab:swift_xrt_log} 	
 \begin{tabular}{ ccccr } 
 \hline 
 \hline 
 \noalign{\smallskip} 
  Sequence   & MJD  & Start time  (UT)        & End time   (UT)       & Exposure    \\ 
            &      & (yyyy-mm-dd hh:mm:ss)   & (yyyy-mm-dd hh:mm:ss)  &(s)        \\
 \noalign{\smallskip} 
 \hline 
 \noalign{\smallskip} 
00016062001	&	60094.76896	&	2023-05-30 15:09:17	&	2023-05-30 21:45:17	&	5661	\\
00016062002	&	60097.67922	&	2023-06-02 05:07:15	&	2023-06-03 03:28:53	&	4418	\\
00016062003	&	60098.51066	&	2023-06-03 08:05:48	&	2023-06-03 16:24:53	&	4403	\\
00016062004	&	60103.59945	&	2023-06-08 13:30:31	&	2023-06-08 15:15:53	&	2146	\\
00016062005	&	60104.29692	&	2023-06-09 03:49:18	&	2023-06-09 10:22:53	&	1647	\\
00016062006	&	60106.25109	&	2023-06-11 05:01:17	&	2023-06-11 07:01:52	&	2119	\\
00016062007	&	60108.36176	&	2023-06-13 07:47:57	&	2023-06-13 09:33:54	&	2131	\\
00016062008	&	60110.28413	&	2023-06-15 05:52:22	&	2023-06-15 07:45:54	&	2397	\\
00016062009	&	60113.83124	&	2023-06-18 19:53:03	&	2023-06-18 20:00:55	&	471	\\
00016062010	&	60120.64475	&	2023-06-25 12:11:59	&	2023-06-25 18:44:52	&	1918	\\
00016062011	&	60123.61480	&	2023-06-28 12:55:43	&	2023-06-28 16:34:53	&	2480	\\
00016062012	&	60126.45835	&	2023-07-01 09:15:10	&	2023-07-01 12:44:52	&	2517	\\
00016062013	&	60129.53083	&	2023-07-04 05:24:28	&	2023-07-04 20:04:19	&	2379	\\
  \noalign{\smallskip}
  \hline
  \end{tabular}
  \end{center}
  %\begin{list}{}{} 
  %\item[$^{\mathrm{a}}$] 
  %\end{list} 
%\end{table*} 
  \end{table} 

\end{appendix}
\end{document}